\documentclass[aps,apl,twocolumn,showpacs]{revtex4}
\usepackage{graphicx}
\usepackage{hyperref}

\begin{document}

\title{Spin-polarized quasiparticle tunneling in spin-filter pseudospin-valve devices}
\author{P. K. Muduli}\email{muduli.ps@gmail.com}

\affiliation{Department of Materials Science and
Metallurgy,University of Cambridge, 27 Charles Babbage Road,
Cambridge CB3 0FS,United Kingdom}

\date{\today}
\begin{abstract}
Spin selective nature of spin-filter tunnel junctions can be
integrated with conventional metallic ferromagnets to regulate
spin polarized quasiparticles in superconducting devices. We
report fabrication of pseudo spin-valve device made with a bilayer
of nitride spin-filter tunnel barrier (DyN or GdN) and a
transition metal ferromagnet (Co and Gd). We show resistance
switching in these devices corresponding to parallel and
antiparallel configuration of their mutual magnetization
direction. With optimal deposition process partial nitridation of
the Co layer can be achieved. The magnetically dead native CoN$_x$
layer at the Co-DyN interface acts the role of the barrier in
these devices. In pseudo spin-valve with Co, lower resistance was
found for antiparallel state compared to parallel configuration.
Reverse resistance switching behavior was observed for the pseudo
spin-valves with Gd. Presence of resistance switching in these
devices further confirm the spin-filtering nature of DyN and GdN
tunnel barrier. Quasiparticle transport at different temperatures
in these devices was found to be compatible with conventional
N-I-S tunnelling model. These devices can be further engineered to
regulate spin polarized supercurrent in superconducting
spintronics devices.

\end{abstract}
\maketitle
\clearpage
\section{Introduction}
Superconducting spintronics depends on the creation and
manipulation of spin polarized current in devices involving
superconducting (S) and ferromagnetic (F)
materials\cite{Linder,Blamire,Wakamura}. In particular, Josephson
junction of the S-F-S type have got a lot of attention  recently
due to their potential application in quantum computing and
spintronics \cite{Makhlin,Clarke,Yamashita}. The spin singlet
Cooper pair amplitude undergoes an oscillatory decay inside a
ferromagnetic metal. This oscillatory behavior can lead to a phase
difference of $\pi$ between the two superconductors depending on
the thickness of the ferromagnet in the S-F-S
junction\cite{Buzdin,Ryazanov,Kontos}. The possibility of $\pi$
junction was initially  proposed by Bulaevskii \emph{et. al.} for
a Josephson junction including magnetic impurities in the tunnel
barrier\cite{bulaevskii}. A lot of experimental and theoretical
investigation has been done on $\pi$ Josephson junctions since
then\cite{Bell,jason-prl,Sellier,Halterman}. In the case of tunnel
junctions the transition from 0 to $\pi$-state can be
distinguished in the differential conductance spectra in the
quasiparticle tunnelling regime. In Al-Al$_2$O$_3$-PdNi-Nb tunnel
junctions when thickness of the ferromagnetic layer is increased
features in the superconducting density of state (DOS) are
reversed with respect to the normal state indicating a 0 to $\pi$
state transition\cite{Kontos}. More complicated tunnel junctions
of the from SFIFS\cite{krivoruchko,bergeret},
SIFIS\cite{Petkovic}, SIFS\cite{wides,larkin,Wild} and
SIsFS\cite{bakurskiy,Bakurskiy-prb,vernik,ryazanov} has also been
proposed theoretically and observed experimentally. However,
compared to metallic S-F-S Josephson junction tunnelling devices
have been relatively poorly explored experimentally sofar. It has
been theoretically predicted that it is possible to generate
spin-triplet superconducting correlations in a ferromagnet with
magnetic inhomogeneity in contact with a
superconductor\cite{bergeret,eschrig}. The triplet correlations
decay over much longer length scale than usual. Some experimental
evidences has also been reported for their
existence\cite{sosnin,robinson-science,khaire,keizer,anwar,jwang}.
However, a democratic experimental evidence of long-range
odd-frequency spin-triplet pairs is still missing. As tunnelling
process is more spin selective than the diffusive counter part,
spin-valves involving tunnel barriers can provide more definitive
evidence for spin-triplet pairs

Recently, we have shown that magnetic semiconductors like GdN,
DyN, etc., are quite compatible with superconducting NbN and are
very promising materials for superconducting
spintronics\cite{muduli,muduli-arxive,pal,blamire,senapati}. We
have shown that spin polarized tunnel current can be very
effectively generated due to spin filtering through these tunnel
barriers. With unique properties of spin-filter tunnel junctions
it is possible to create composite structures in combination with
ferromagnets to design new kind of superconduting spintronics
devices\cite{Nagahama,baek,leclair,moodera-rev,shiro,Giazotto,Kawabata}.
However, integrating normal ferromagnets within all nitride device
is not trivial and possible nitridation of the metal layers and
nature of interfaces have to be carefully considered. In this
paper, we report fabrication and electrical characterization of
psuedospin-valve devices made of a strong ferromagnet and a
spin-filter tunnel barrier. We have used GdN and DyN tunnel
barrier as spin polarizer and Co and Gd as analyzer. A comparison
of resistance switching behaviour has been done between Co-DyN and
Gd-GdN type devices. Devices with different thickness of DyN were
fabricated to optimize tunnelling regime. The tunneling nature of
the device has been analyzed through current-voltage (I-V)
measurements at different temperatures. Quasiparticle transport
through the psuedospin-valve devices has been compared to N-I-S
tunnelling model. We show spin regulation can be achieved in these
devices by controlling relative magnetization of the two magnetic
layers. We provide a  qualitatively explanation for  the
resistance switching behavior observed in our psuedo spin-valve
device.

\section{Experimental}
Multilayer structures NbN-FM-FI-NbN were fabricated by DC
sputtering in an ultrahigh vacuum (UHV) chamber at room
temperature (here FM = Co, Gd and FI= DyN, GdN). Nitride layers
were deposited by sputtering of high purity metal targets in an
Ar/N$_2$ gas mixture with deposition condition as described in the
reference\cite{muduli,muduli-arxive}. The Co layer was deposited
at 1.5 Pa in a pure Ar gas with 40 W sputtering power.
Ferromagnetic Gd was also deposited in a similar condition with
lower 20 W sputtering power. Thickness of different layers were
controlled by regulating the speed of a rotating substrate stage.
The multilayer structure NbN-FM-FI-NbN was deposited in the order
from left to right insitu without breaking vacuum. The nitridation
of Co and Gd during subsequent deposition of DyN or GdN strongly
depends on time they are exposed to the nitride plasma. Therefore,
thickness of the DyN or GdN layer plays a crucial role in
controlling the interfacial nitridation. In thicker films the Co
(Gd) layer is exposed to nitride plasma for a longer time which
can lead to higher nitridation. The CoN$_x$ thin film is known to
be magnetic with a perpendicular magnetic
anisotropy\cite{Matsuoka}. However, a very thin layer of CoN$_x$
might be magnetically dead and prevent magnetic coupling between
the FI and FM. In the case of multilayer with Gd, nitrogen
deficient GdN$_x$ might be formed at the interface. The thickness
of the top and bottom NbN was kept fixed at 50 nm in all the
multilayers. While thickness of the Co and DyN layer was varied in
the series. As the top and bottom NbN is common in all the
devices, we have used abbreviation FM($t_{FM}$ nm)-FI($t_{FI}$ nm)
throughout this paper to represent different kind of devices. Here
$t_{FM}$ and $t_{FI}$ represent thickness of the ferromagnetic
metal and insulator in nanometer, respectively. Devices were
fabricated using photolithography in a mesa structure. The
junction dimension was defined by etching the top NbN in CF$_4$
plasma for 30 sec and milling 10 min with Ar-ion after that. Top
contact was made with Nb electrode after SiO$_2$ lift-off. Lateral
dimension of the junctions were 7 $\mu$m $\times$ 7 $\mu$m.
Schematic of the mesa device is shown in the inset of Fig. 1.
Differential conductance ($dI/dV$) was measured with a lock-in
technique in a custom made dip-stick using liquid helium. Spin
valve measurements were done in a closed-cycle helium refrigerator
from \emph{Cryogenics Lmt.} The measurements were done in a
four-probe configuration with DC bias current. The magnetization
of the multilayer films deposited at the same time were carried
out with a SQUID magnetometer. In this paper measurements done on
one representative sample is shown. The reproducibility and
behaviour of other devices are shown in the supplementary
material.

\section{Results and Discussion}

\begin{figure}[!h]
\begin{center}
\abovecaptionskip -10cm
\includegraphics [width=8 cm]{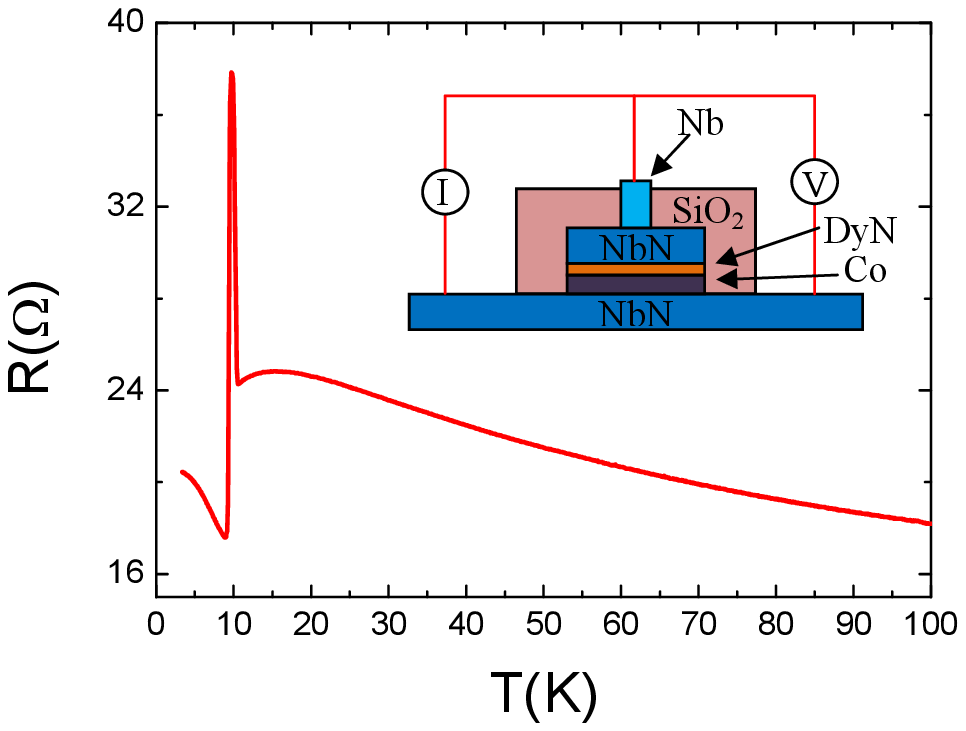}
\end{center}
\caption{\label{fig1} (Color online) Temperature dependence of
resistance of a Co(5 nm)-DyN(2 nm) device. The measurement was
done using a current $I$ = 100 $\mu$A. Inset shows schematic of
the mesa structure used for devices. }
\end{figure}
Figure 1 shows temperature dependence of resistance $R(T)$ of a
Co(5 nm)-DyN(2.5 nm) device measured with a bias current I = 100
$\mu$A. The resistance showed semiconducting temperature
dependence with a small deviation below 20 K. Similar $R(T)$ was
also found for Co-GdN devices (see supplementary figure SFig. 2).
The temperature dependence of resistance in these devices is most
likely determined by the most resistive part i.e., Co-DyN
interface, where formation of disordered CoN$_x$ is possible. See
supplementary material (SFig. 3) for $R(T)$ of devices with
different thickness of Co and DyN. Superconducting transition of
NbN in the Co(5 nm)-DyN(2.5 nm) device can be seen at $T_C$= 10.6
K as a sharp drop in the resistance. The superconducting coherence
length of NbN in the dirty limit can be determined from the
expression;$ \xi _{NbN}  = (\hbar D_{NbN} /2\pi k_B T_C )^{1/2}$ .
Using diffusion constant $D_{NbN}$= 1.48 $\times$
10$^{-4}$m$^2$s$^{-1}$ and $T_C$=10.6 K we found $\xi _{NbN}$ =
4.1 nm (See Supplementary Information for calculation of diffusion
constant) \cite{kartikapl}. The superconducting coherence length
inside Co can be calculated from the expression; $ \xi _{Co}  =
(\hbar D_{Co} /k_B T_{Curie} )^{1/2}$ ; where $D_{Co}$  and
$T_{Curie}$ are the diffusion constant and the Curie temperature
of Co, respectively. With $D_{Co}$ = 6 $\times$
10$^{-4}$m$^{2}$s$^{-1}$ and $T_{Curie}$ = 1388 K, we found $ \xi
_{Co}$=1.8 nm \cite{jwang}. The thickness of the Co in all our
Co-DyN devices is $\sim$5 nm which is larger than both $\xi _{Co}$
and $\xi _{NbN}$. For conventional spin-singlet case the
superconducting correlations decay over a length $\xi _{Co}$ in
the diffusive ferromagnet. Therefore, our NbN-FM-FI-NbN devices
can be considered as N-I-S type device instead of a S-N-I-S type
device.

\subsection{Tunneling behavior}
\begin{figure}[!h]
\begin{center}
\abovecaptionskip -10cm
\includegraphics [width=8 cm]{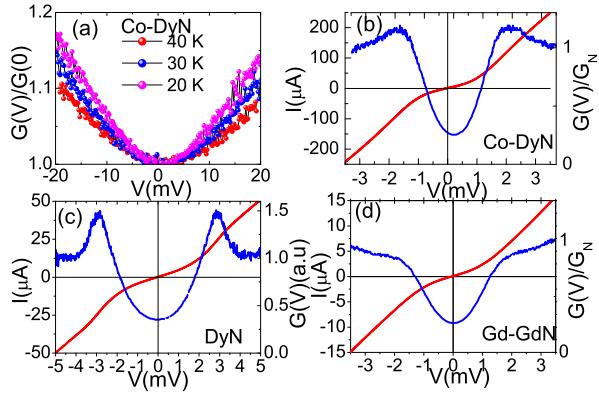}
\end{center}
\caption{\label{fig1} (Color online) (a) $G(V) = dI/dV$
conductance spectra of a Co(5 nm)-DyN(2 nm) tunnel junction
measured at 20, 30 and 40 K. The conductance spectra is normalized
to  $G(0)$ (b) The IV and Normalized conductance spectra of the
same Co(5 nm)-DyN(2 nm) tunnel junction measured at 4.2 K. The
conductance spectra is normalized to normal state conductance
$G_N$. (c) IV and normalized conductance spectra of a NbN-DyN-NbN
tunnel junction measured at 4.2 K. (d)The IV and normalized
conductance spectra of a Gd(5)-GdN(4.5) tunnel junction measured
at 4.2 K.}
\end{figure}
Current-voltage (I-V) measurements were done at different
temperatures to understand the nature of the electrical transport
in the devices. Figure 2(a) shows the conductance $G(V)(= dI/dV)$
normalized to its value at $V = 0$ for the Co(5 nm)-DyN (2.5 nm)
device. Clear parabolic conductance spectra at 20, 30 and 40 K
suggest tunnelling-type transport in this devices through all
temperature range. The $dI/dV$ spectra  of the devices were also
measured at different temperatures below the $T_C$ of NbN. The I-V
and normalized conductance spectra $G(V)/G_N$ of the same junction
measured at 4.2 K is shown in Fig. 2(b). Fully developed
superconducting gap structure with 2$\Delta$ $\sim$2.94 meV can be
seen. For comparison Fig. 2(c) shows I-V and normalized
conductance spectra $G(V)/G_N$ of a NbN-DyN-NbN tunnel junction
without Co. A superconducting gap 4$\Delta$ $\sim$5.68  meV can be
observed in this device. Figure 2(d) shows I-V and normalized
conductance spectra $G(V)/G_N$ of a Gd(5 nm)-GdN(3 nm) device. The
Superconducting gap was found to be 2$\Delta$ $\sim$3.18 meV in
this case.  In the Gd(5 nm)-GdN(3 nm) device the actual thickness
of GdN can be slightly larger than the deposited value due to
partial nitridation of Gd during deposition. As 4 nm GdN is at the
limit of tunnelling to diffusive transport, the gap edges are
slightly smeared in the Gd(5 nm)-GdN(3 nm) device. We have
previously observed that the superconducting gap structure
disappers in the conductance spectra for junctions with DyN
thickness $>$ 4 nm \cite{muduli}. In the Co-DyN devices well
defined gap structure was absent for DyN thickness $>$ 4 nm (see
supplementary figure SFig. 8 for conductance spectra of Co-DyN
devices with different thickness of DyN). The superconducting gap
value $\Delta$ found from the conductance spectra for our
S-FN-FI-S device is slightly greater than the value found for
S-FI-S type devices. Therefore, the Co-DyN and Gd-GdN devices are
not truly N-I-S type device as expected.

To further confirm this the $dI/dV$ spectra below the $T_C$ of NbN
in our devices can be compared with NIS-type tunnel model.
Normalized tunneling conductance of a NIS junction at a bias
voltage $V$ can be written as:
\begin{equation}
\frac{{G_s (V)}}{{G_N (V)}} = \frac{d}{{d(eV)}}\int\limits_{ -
\infty }^\infty  {N(E)} [f(E) - f(E + eV)]dE,
\end{equation}
where $f(E)$ is Fermi-Dirac distribution function and $N(E)$ is
the normalized BCS density of state of the superconductor. Here
$G_N(V)$ is the normal state conductance of the junction.
Following Dynes approach\cite{dynes} the quasiparticle density of
states can be written as, $ N(E) = N(0)\left| {{\mathop{\rm
Re}\nolimits} \left( {\frac{{E/\Delta - i\Gamma }}{{\sqrt
{(E/\Delta  - i\Gamma )^2  - 1} }}} \right)} \right|$. Here the
smearing parameter $\Gamma$ is included to consider finite
lifetime of quasiparticles. Quasiparticles have finite lifetime at
non zero temperature due to presence of some energy levels within
the superconducting gap. Besides, impurities and pinholes in the
tunnel barrier has also shown to contribute to
$\Gamma$\cite{kalpwik,Schrieffer}. In our case magnetism of the
tunnel barrier (FI) and normal electrode (FN) might further add to
$\Gamma$.

\begin{figure}[!h]
\begin{center}
\abovecaptionskip -10cm
\includegraphics [width=8 cm]{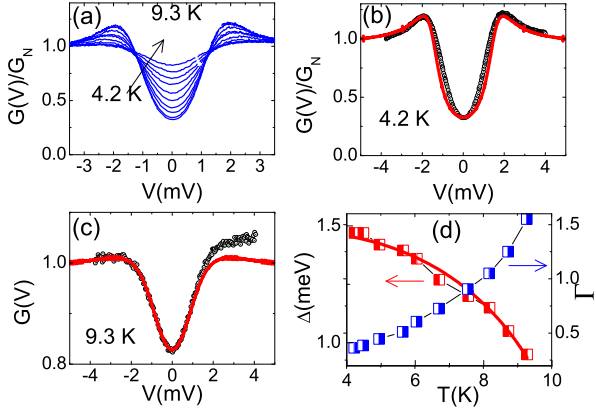}
\end{center}
\caption{\label{fig1} (Color online)(a) Normalized conductance
spectra $G(V)/G_N$ of the Co(5 nm)-DyN(2 nm) tunnel device
measured in the temperature range 4.2 to 9.3 K. (b)NIS-tunneling
model fitting of the conductance spectra measured at 4.2 K.
(c)NIS-tunneling model fitting of the conductance spectra measured
at 9.3 K (d)Temperature dependence of superconducting gap $\Delta$
and smearing parameter $\Gamma$ obtained from fitting. The red
solid line shows fitting to the BCS temperature dependence of
$\Delta(T)$. }
\end{figure}

Figure 3(a) shows the temperature evolution of conductance spectra
of the Co(5 nm)-DyN(2.5 nm) device measured in the range 4.2 to
9.3 K. Fitting of the Eq. (1) (red solid line) to the conductance
spectra measured at 4.2 and 9.3 K is shown in Fig. 3(b) and (c),
respectively. We found acceptable fit of the conductance spectra
to the NIS-tunneling model at all  temperatures by adjusting
smearing parameter $\Gamma$. Fig. 3(d) shows plot of extracted
fitting parameters $\Gamma$ and $\Delta$ at different temperature.
The red solid line shows a typical BCS type temperature
dependence\cite{Takayanagi}:  $ \Delta (T) = \Delta (0)\tanh
(1.74\sqrt {(T_C  - T)/T} ) $, with 2$\Delta (0)$ = 2.97 meV and
$T_C$ = 10.94 K.  The increase in subgap conductance might be due
to magnon-assisted Andreev reflection process which can provide
additional channel for subgap transport\cite{Tkachov}.
Electromagnetic fluctuation in the environment can also lead to
similar situations\cite{pekola,Marco}. Moreover, we could not
observe any feature related to 0 and $\pi$ transition due to poor
coupling between the two NbN layers.

\subsection{Spin-valve behavior}

Figure 4(a) shows magnetic field dependence of resistance of the
Co(5 nm)-DyN(2 nm) device measured at 2 K. The measurement was
done with a bias current $I = 500 \mu$A. The resistance was found
to increase with magnetic field up to $H_p$ $\sim$ 0.1 T and
decrease afterwards. The value of $H_p$ was found to vary from
device to device and a complete decreasing trend was observed for
devices with thicker DyN (see Supplementary figure SFig. 6). Fig.
4(b) shows field dependence of magnetization measured at 12 and
100 K of a NbN-Co(5 nm)-DyN(4.5 nm)-NbN film deposited at the same
time. As 4.5 nm thick DyN have very small magnetic moment compared
to Co the M-H loop is mostly dominated by Co. Fig. 4(c) shows
field dependence of resistance in the field range $\pm$30 mT. One
can clearly see resistance switching at $\pm$25 mT. This switching
appears due to switching of the Co magnetization at $\pm$25 mT.
Although, coercive field of the NbN-Co(5 nm)-DyN(4.5 nm)-NbN film
is 7 mT, resistance switching at 25 mT in the Co(5 nm)-DyN(2 nm)
device is due to the reduced dimension of the junction i.e., 7
$\mu$m $\times$ 7 $\mu$m. We found slight variation in the
switching field from device to device depending on the thickness
of DyN (See supplementary figure SFig. 5). This might be due to
the different nitridation of Co which makes effective Co layer
thinner. One striking feature to notice is a low resistance state
for antiparallel configuration in comparison to parallel
configuration. Fig. 4(d) shows similar measurements on Gd(5
nm)-GdN(3 nm) device. A higher resistance for antiparallel state
can be seen in this case. In Gd-GdN devices resistance switching
was found only at low temperature below 13 K (see supplementary
figure SFig.7). This might be due to the absence of a well defined
magnetic decoupling layer at the interface between Gd and GdN. In
this case a gradient of nitrogen deficiency might separate Gd and
GdN layer from each other providing poor magnetic isolator.

\begin{figure}[!h]
\begin{center}
\abovecaptionskip -10cm
\includegraphics [width=8 cm]{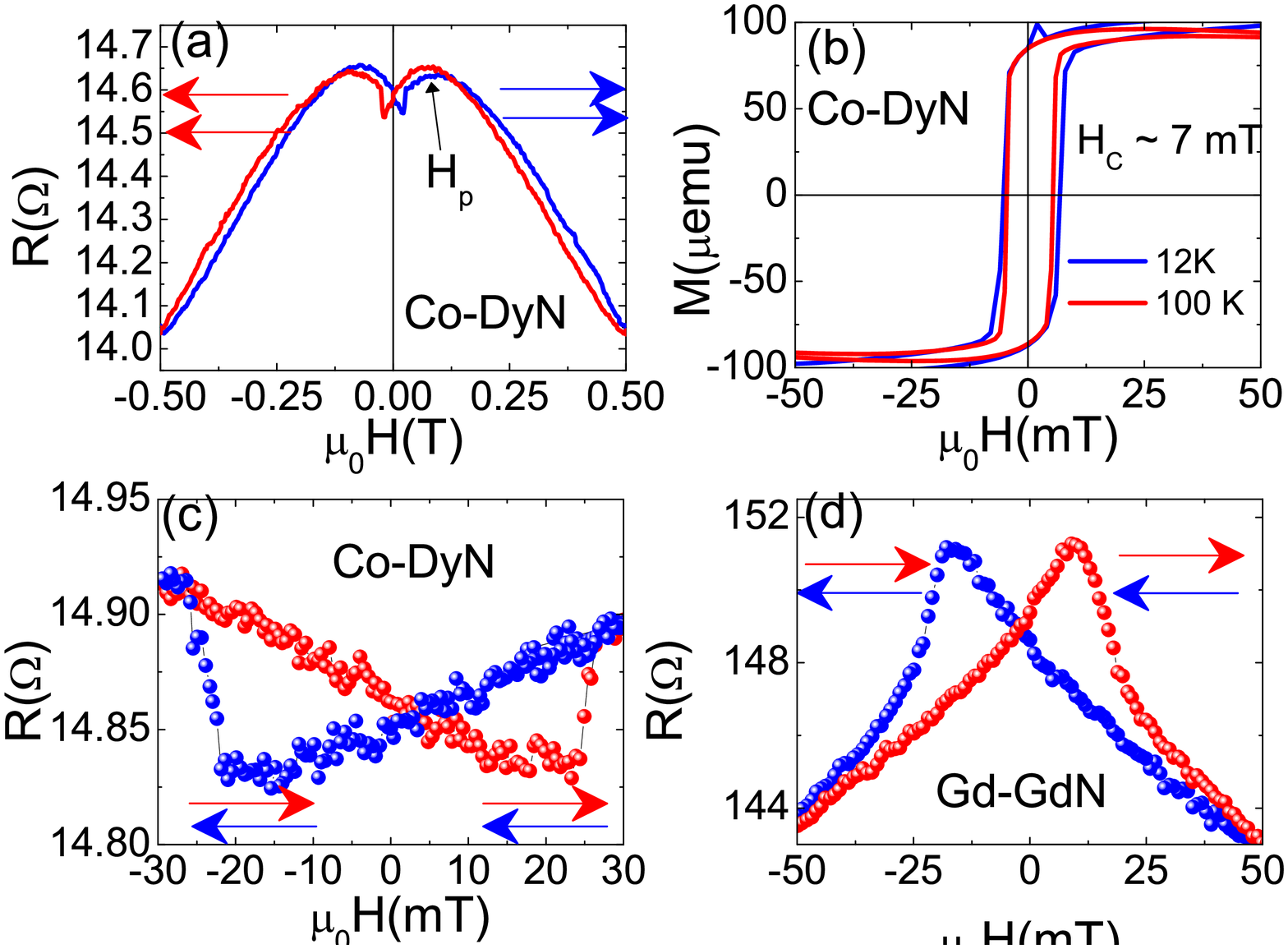}
\end{center}
\caption{\label{fig1} (Color online)(a) Magnetic field dependence
of resistance of a  Co(5 nm)-DyN(2 nm) device measured at 2 K
using a current $I$ = 500 $\mu$A. (b) Magnetic field dependence of
magnetization at 12, 20, and 100 K for a NbN-Co(5 nm)-DyN(4.5
nm)-NbN film deposited at the same time. (c)R-H loop of the Co(5
nm)-DyN(2 nm) device in the low-field range. (d)Field dependence
of resistance of a  Gd(5 nm)-GdN(4.5 nm) device measured at 2 K.}
\end{figure}

Figure 5  shows R-H loops of the Co(5 nm)-DyN(2 nm) device
measured at different temperatures. The resistance switching was
found to disappear as temperature was increased to 30 K. This is
expected as DyN gets into paramagnetic phase above $T_{Curie}$
$\sim$35 K. Temperature dependence of switching strongly suggest
that the switching arises due to the relative mutual magnetization
of Co and DyN. We could not find any switching in devices with
thicker DyN ($>$ 3.5 nm). This advocate other sources of switching
like AMR of Co or artefact due to device geometry is not
responsible for such resistance switching. In samples with thicker
DyN the interfacial CoN$_x$ is most likely very thick which causes
loss of spin polarization of electrons after filtering through
DyN. The high field magnetoresistance seen in Figure 4(a,d) seems
likely to originate from a field-enhanced magnetisation
\cite{muduli-arxive} and hence an exchange splitting which
increases with field.

\begin{figure}[!h]
\begin{center}
\abovecaptionskip -10cm
\includegraphics [width=8 cm]{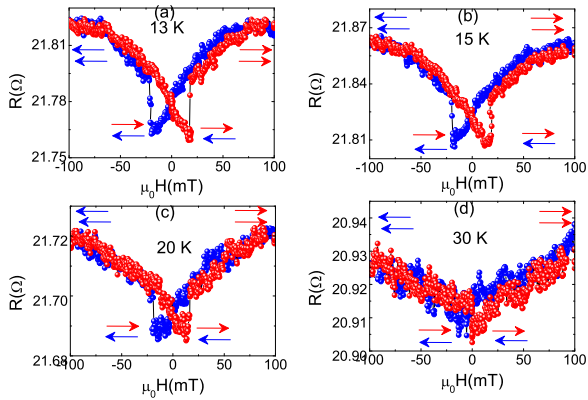}
\end{center}
\caption{\label{fig1} (Color online)Field dependence of resistance
of the NbN-Co(5 nm)-DyN(2 nm)-NbN  device measured at (a) 13 K (b)
15 K, (c)20 K  and (d) 30 K. Measurements were done with a current
$I = 500 \mu A$.}
\end{figure}

\begin{figure}[h]
\begin{center}
\abovecaptionskip -10cm
\includegraphics [width=8 cm]{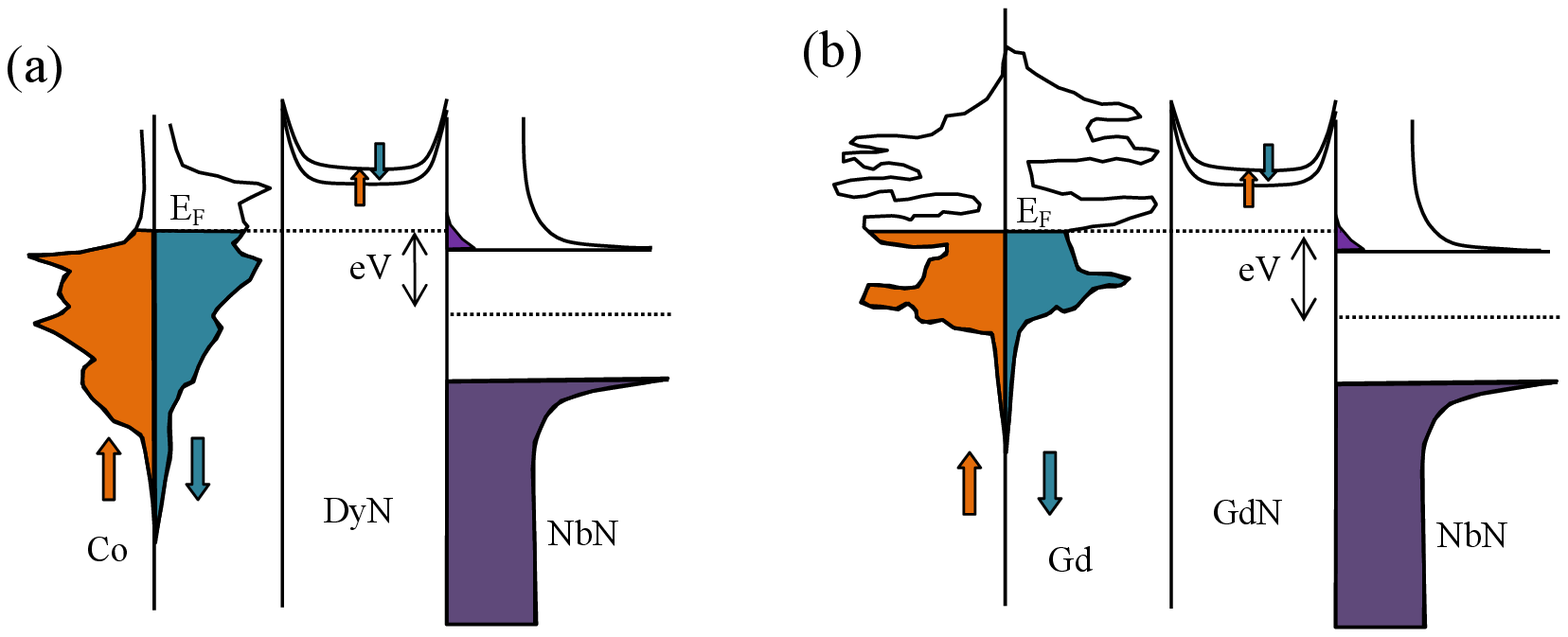}
\end{center}
\caption{\label{fig1} (Color online) Schematic of the
spin-resolved density of state for the (a)Co-DyN and (b)Gd-GdN
device. The density of state of Co and Gd is shifted by $eV$ along
the energy axis when a bias voltage $V$ is applied. All spin-valve
measurements were done with bias voltage $eV > \Delta$.}
\end{figure}

The resistance switching in our devices can be understood
considering spin dependent density of states (DOS) of different
layers as shown in Fig. 6. When magnetization of Co and DyN are
parallel to each other, the up-spin electrons tunneling through
DyN experience a lower barrier height compared to down-spin
electrons. During tunneling process spin orientation of electrons
are conserved, therefore, up and down-spin electrons can tunnel
only into spin-up and spin-down  states of Co,
respectively\cite{jullier}. Similarly when magnetization of Co and
DyN are antiparallel to each other down-spin electrons are
filtered through. In the case of Co the DOS  at the Fermi level
$E_F$ for up-spin electrons is lower than DOS for down-spin
electrons\cite{coey}. Therefore the transparency through the
DyN-Co bilayer will be higher for antiparallel configuration
compared to parallel configuration. This will cause a
low-resistance state for situation when magnetization of Co and
DyN are antiparallel to each other. This concept can be further
verified in Gd-GdN devices where Co is replaced by Gd. Schematic
of spin dependent DOS of Gd is shown in Fig. 6(b). In case of Gd
the up-spin DOS at $E_F$ is slightly higher than down-spin
DOS\cite{Harmon}. Moreover, unlike Co in the case of Gd tunneling
spin polarization is usually found positive
experimentally\cite{Tedrow}. Therefore, one expect a higher
resistance for antiparallel configuration compared to parallel
configuration as observed in normal spin-valves\cite{jullier}. The
resistance switching behavior shown in Fig. 4 confirms our
interpretation. The sign of the tunnel spin polarization is known
to depend strongly on the nature of the interface bonding which is
likely different between Co-DyN and Gd-GdN. A detailed band
structure calculation is needed to understand different sign of
spin polarization in these interfaces.
\section{Conclusions}
In conclusion, we have fabricated NbN-FM-FI-NbN psuedospin-valve
devices (with FM = Co, Gd and FI= DyN, GdN) and made an extensive
study of electrical transport measurements. Tunnelling regime was
achieved in these devices by optimizing thickness of the DyN
layer. In the tunnelling regime quasiparticle tunnelling spectra
through the spin-valve were compared to tunneling spectra of NIS
tunnel model with a nonmagnetic barrier. We also measured R-H loop
of the devices at different temperatures. Clear resistance
switching was observed in these devices corresponding to their
mutual magnetization direction. The resistance switching was found
to be sensitive to density of state (DOS) at the FM-FI interface.
In the case of Co a low resistance state was found for
antiparallel configuration due to lower spin-up DOS at $E_F$. A
reverse behaviour with high resistance for antiparallel
configuration was found for Gd devices. The resistance switching
in these spin valve measurements confirm the spin filtering nature
of DyN tunnel barrier. Optimization of such devices with ideal
deposition condition and interface can lead to huge MR. Thickness
of Co (Gd) and barrier transparency of DyN (GdN) can be further
tuned to create stronger coupling between the two NbN leading to a
S-I-S type device with a spin-valve sandwiched in between.This
kind of devices can be tuned between 0 and $\pi$ state through
magnetization configuration of the spin valve\cite{baek,gingrich}.

\noindent  \textbf{Acknowledgments}\\
This work was done when PKM was supported by the ERC Advanced
Investigator Grant SUPERSPIN during  the period  June 2012-April
2015. PKM acknowledges Dr X. L. Wang for SQUID measurements and
Device Materials Group (DMG) for experimental facility.

\newpage
\setcounter{figure}{0}
\renewcommand{\figurename}{SFig.}
\begin{widetext}

\begin{center}
\textbf{\huge\underline{Supplementary Information}}
\end{center}
\begin{center}
\large\emph{Spin-polarized quasiparticle tunneling in spin-filter
pseudospin-valve devices}
\end{center}

Pseudospin-valve devices were fabricated from NbN-FM-FI-NbN
multilayered thin films  deposited by DC sputtering method. Here
FM = Co and Gd;  FI = DyN and GdN. A series of devices were
fabricated from Co(5 nm)-DyN(\emph{t}) multilayer with different
thickness of DyN. Few representative devices were fabricated in
the configuration Co-GdN and Gd-GdN for comparison. More detailed
experiment was done in the Co-DyN series compared to others. In
this manuscript a comparison has been made between devices with
GdN and DyN for spin-valve measurements. The comparison is
reasonable as GdN and DyN are similar ferromagnetic semiconductors
with only different spin-filtering efficiency. Measurements done
on different devices are summarized in the figures below.

\textbf{Calculation of Diffusion constant of NbN :}

\begin{figure}[!h]
\begin{center}
\abovecaptionskip -10cm
\includegraphics [width=8 cm]{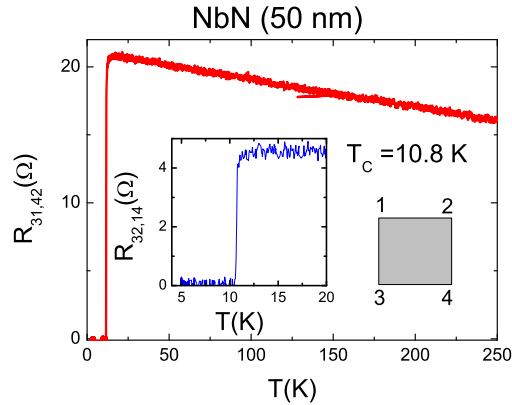}
\end{center}
\caption{\label{fig5}(Color online)Temperature dependence of
resistance of a 50 nm thick NbN film deposited under similar
condition as the devices. Inset shows measurement done in a
different Van der Pauw configuration. Here residual resistivity
RR($R_{RT}/R_{15 K}$) $\approx$0.74. Resistivity determined by Van
der Pauw method to be $\rho(15 K) \approx$ 241.2 $\mu\Omega-cm$.}
\end{figure}

The diffusion constant can be calculated from Einstein relation;
\begin{equation}
D = \frac{1}{{e^2 D(E_f )\rho }}
\end{equation}

Where $ {D(E_f )}$ is the density of state at Fermi energy and
$\rho$ is the resistivity. The diffusion constant can also be
written as; $ D = \frac{1}{3}v_F l_e$, with $v_F$ is Fermi
velocity and $l_e$ is the average electron mean free path. For NbN
with $D(E_f ) = 1.74 \times 10^{28} eV^{ - 1}m^{-3}$ [Chockalingam
\emph{et. al.} Phys. Rev. B \textbf{2008}, 77, 214503.] and $\rho
= 2.41 \mu\Omega-m$, we found  $D = 1.48 \times 10^{-4}
m^2s^{-1}$.

\begin{figure}[!h]
\begin{center}
\abovecaptionskip -10cm
\includegraphics [width=8 cm]{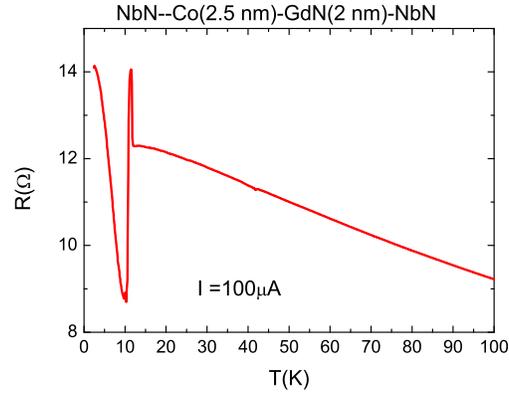}
\end{center}
\caption{\label{fig5}(Color online)Temperature dependence of
resistance of a Co(2.5 nm)-GdN(2 nm) device measured with a
current I =100 $\mu$A. The $R(T)$ in these devices may not be
determined by GdN but more accurately the temperature dependence
is  decided by the composite of Co-GdN. For devices with thinner
Co, complete nitridation of Co cannot be ruled out. Therefore,
$R(T)$ may vary from device to device.}
\end{figure}

\begin{figure}[!h]
\begin{center}
\abovecaptionskip -10cm
\includegraphics [width=8 cm]{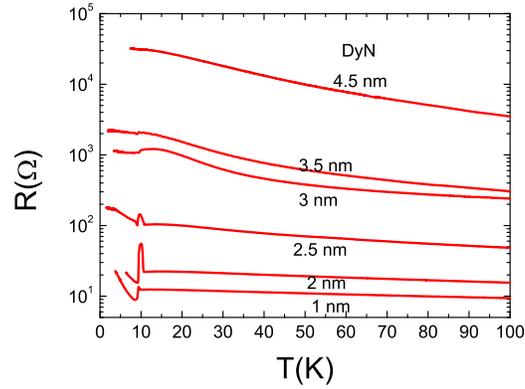}
\end{center}
\caption{\label{fig5}(Color online)Temperature dependence of
resistance of Co(2.5 nm)-DyN(\emph{t} nm) devices with different
thicknesses of DyN.}
\end{figure}

\begin{figure}[!h]
\begin{center}
\abovecaptionskip -10cm
\includegraphics [width=8 cm]{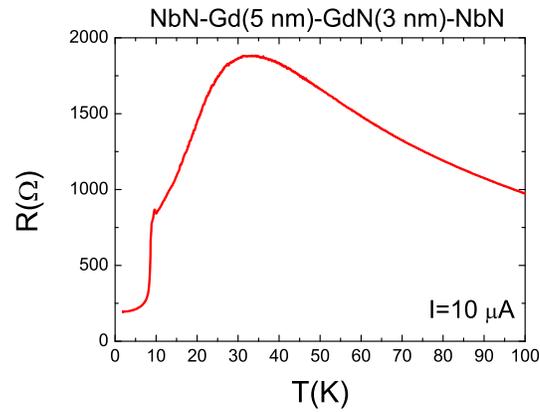}
\end{center}
\caption{\label{fig5}(Color online)Temperature dependence of
resistance of a Gd(5 nm)-GdN(3 nm) device measured with a bias
current I = 10 $\mu$A. }
\end{figure}

\begin{figure}[!h]
\begin{tabular}{lll}

  \centering
  \includegraphics[width= 6 cm]{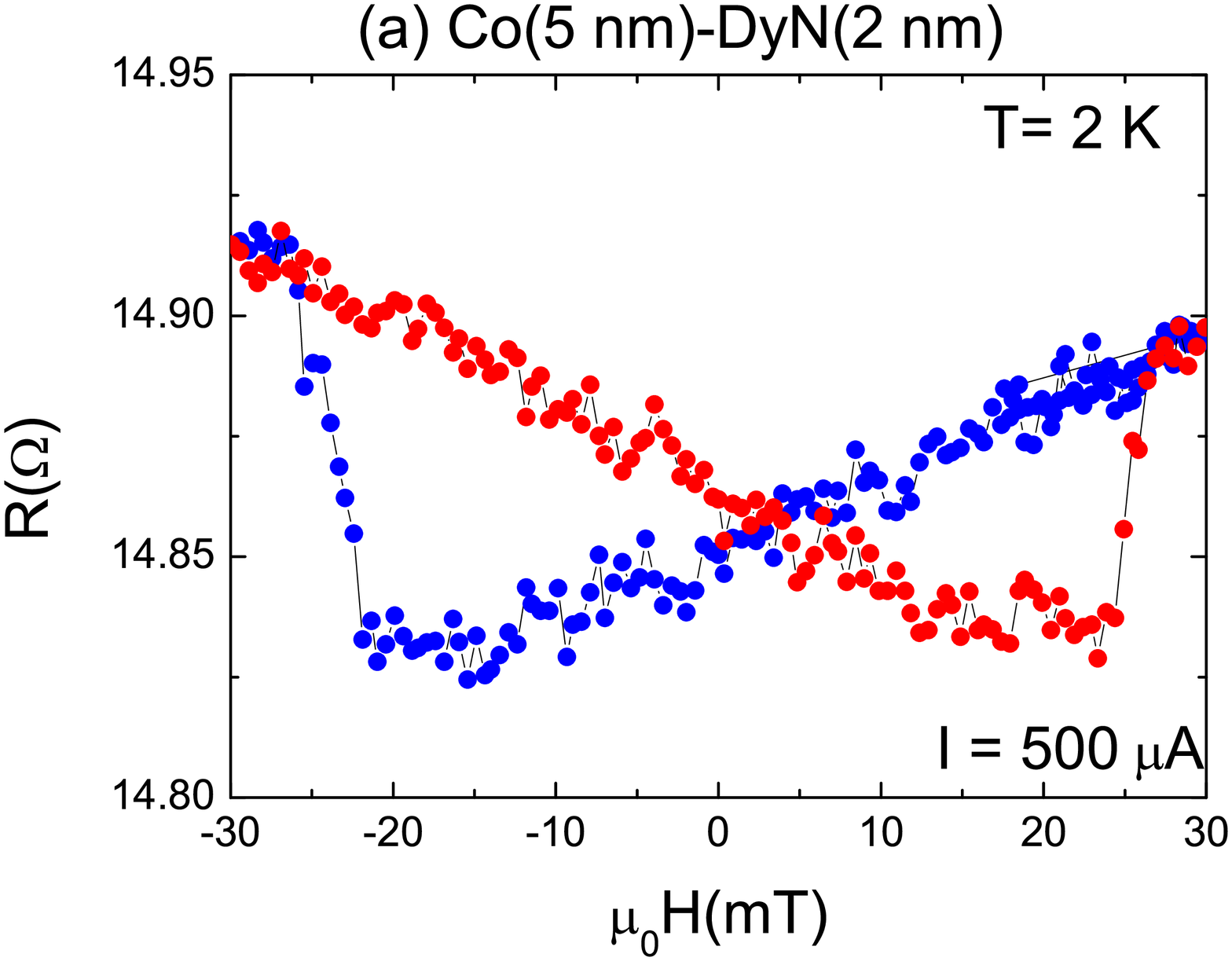}
&

  \includegraphics[width= 6 cm]{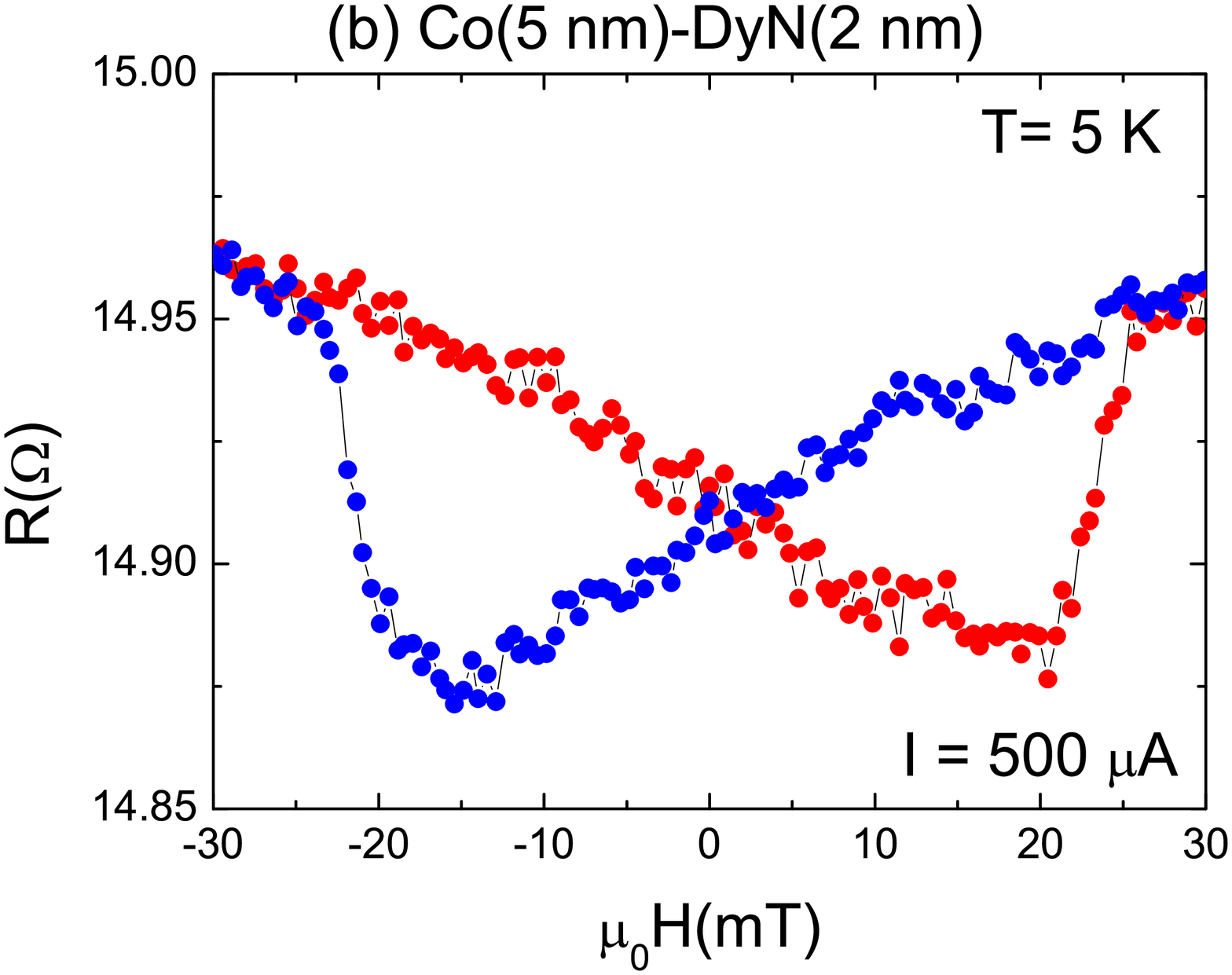}
&

  \includegraphics[width= 6 cm]{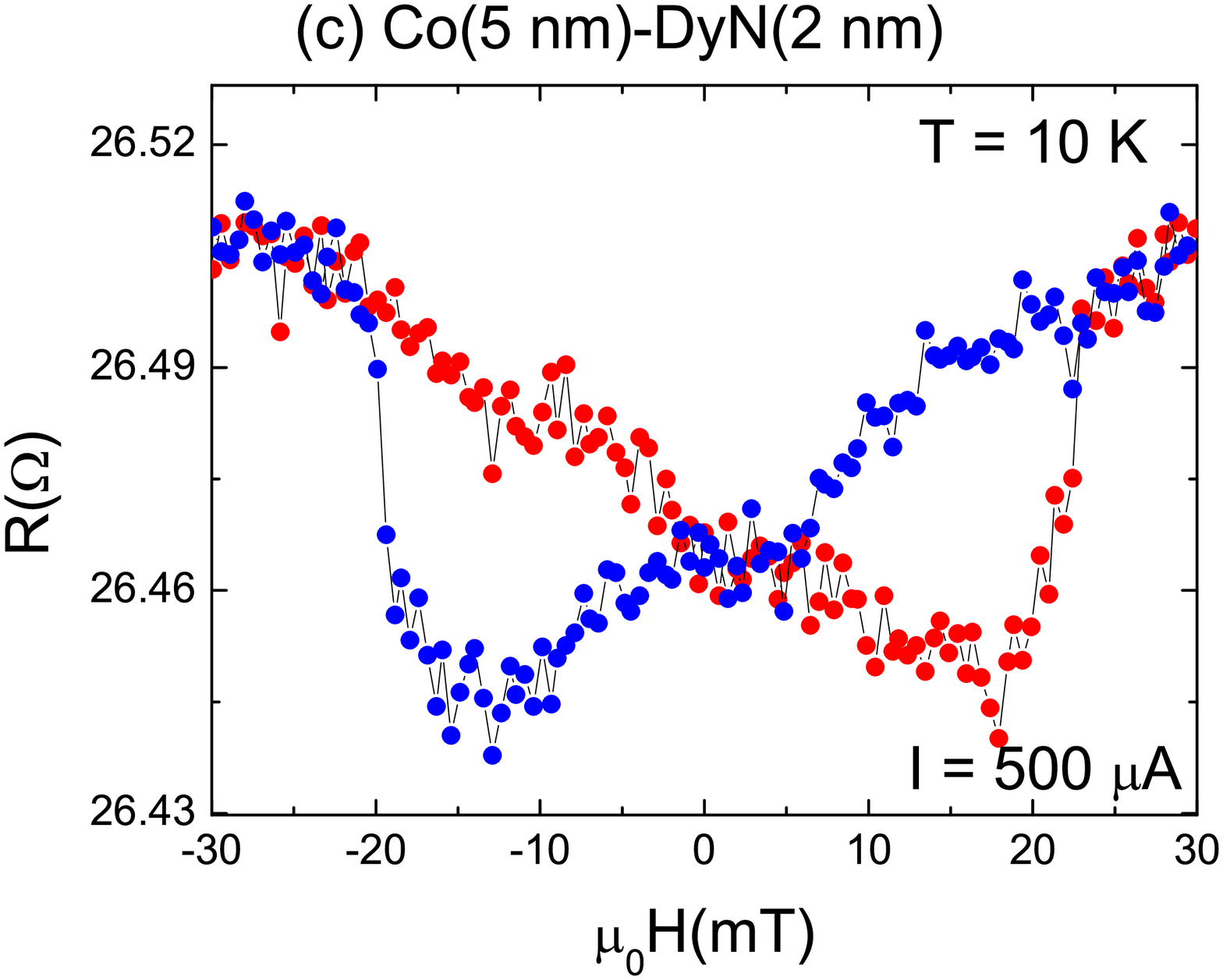}\\

  \centering
  \includegraphics[width= 6 cm]{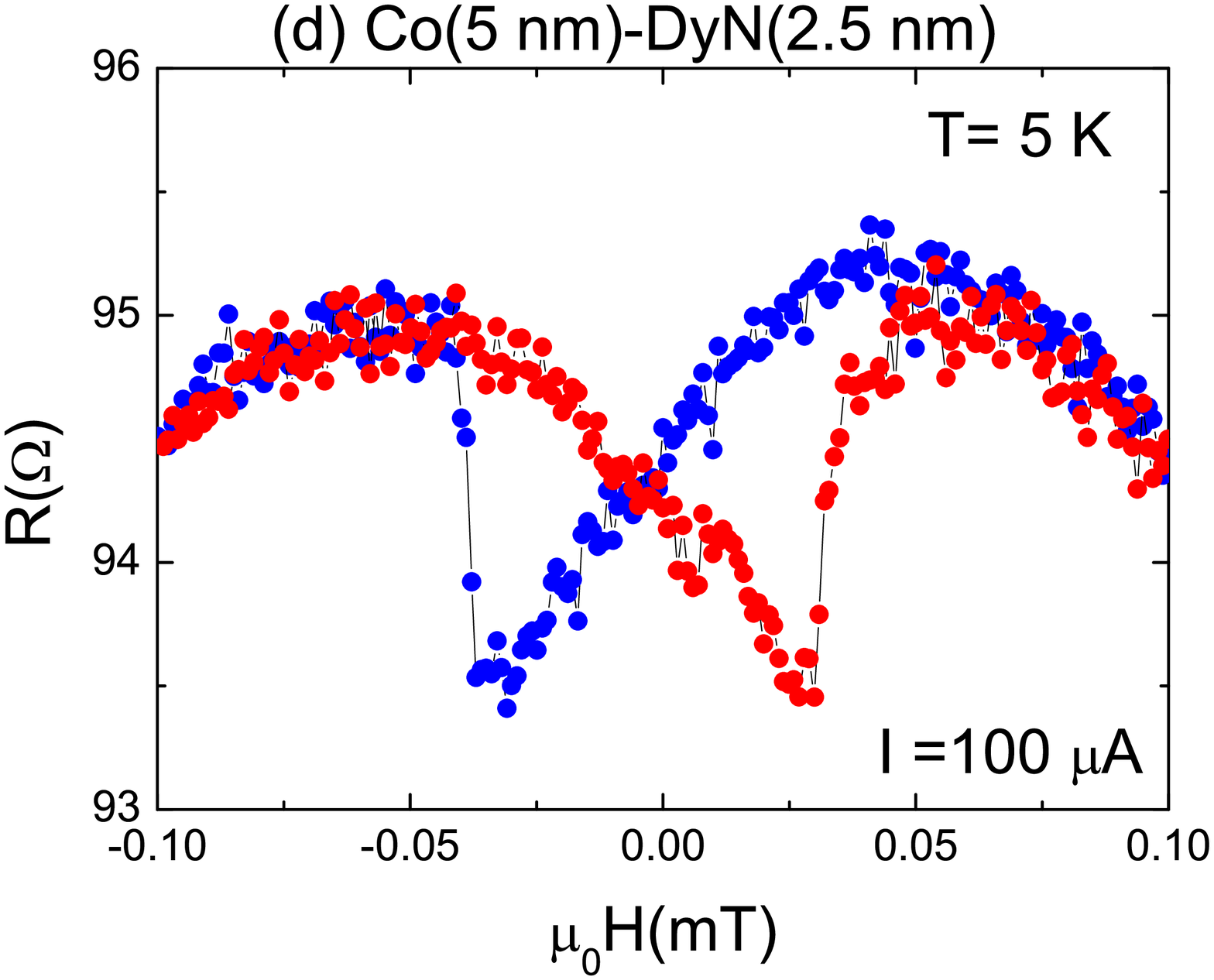}
&

  \includegraphics[width= 6 cm]{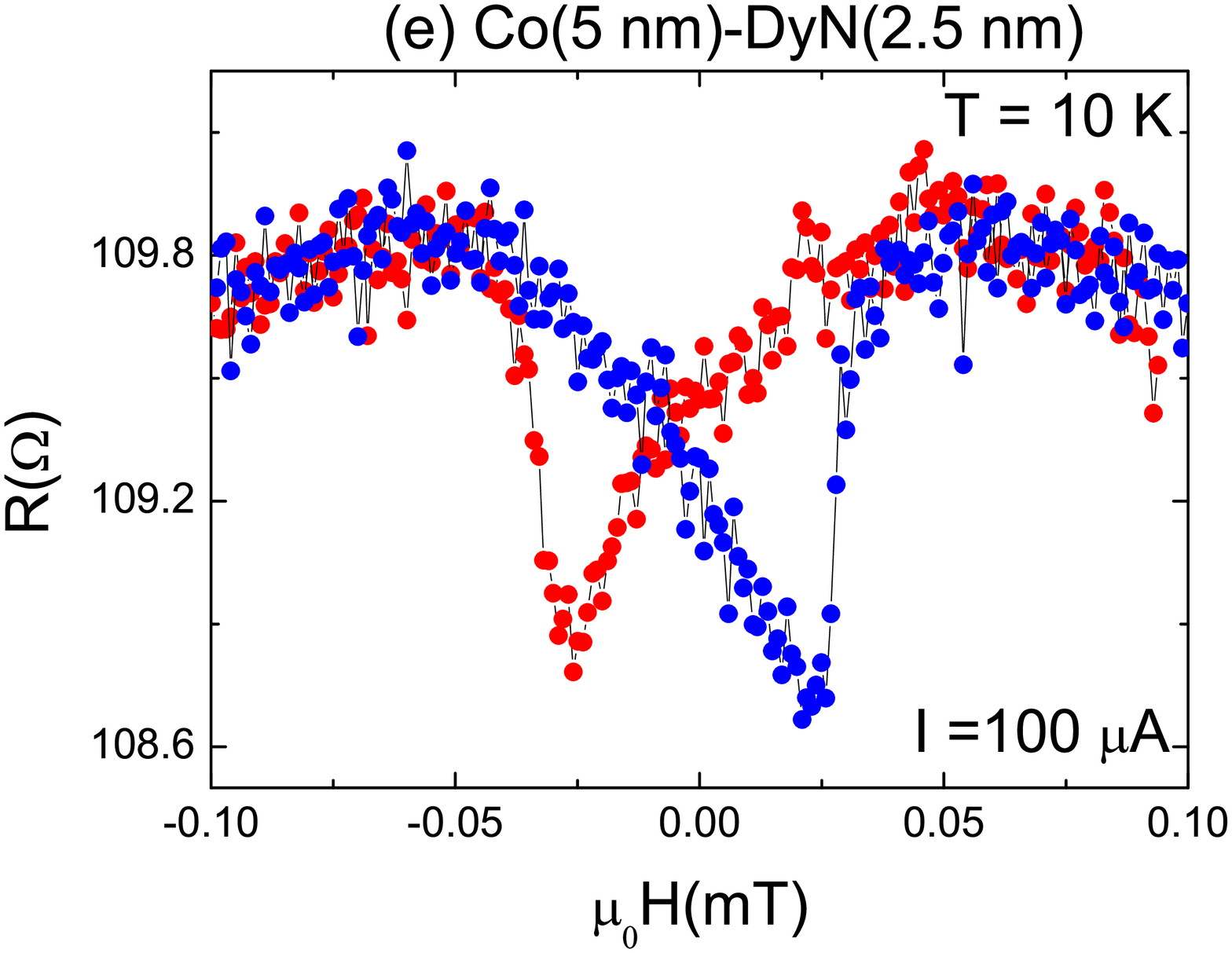}
&

  \includegraphics[width= 6 cm]{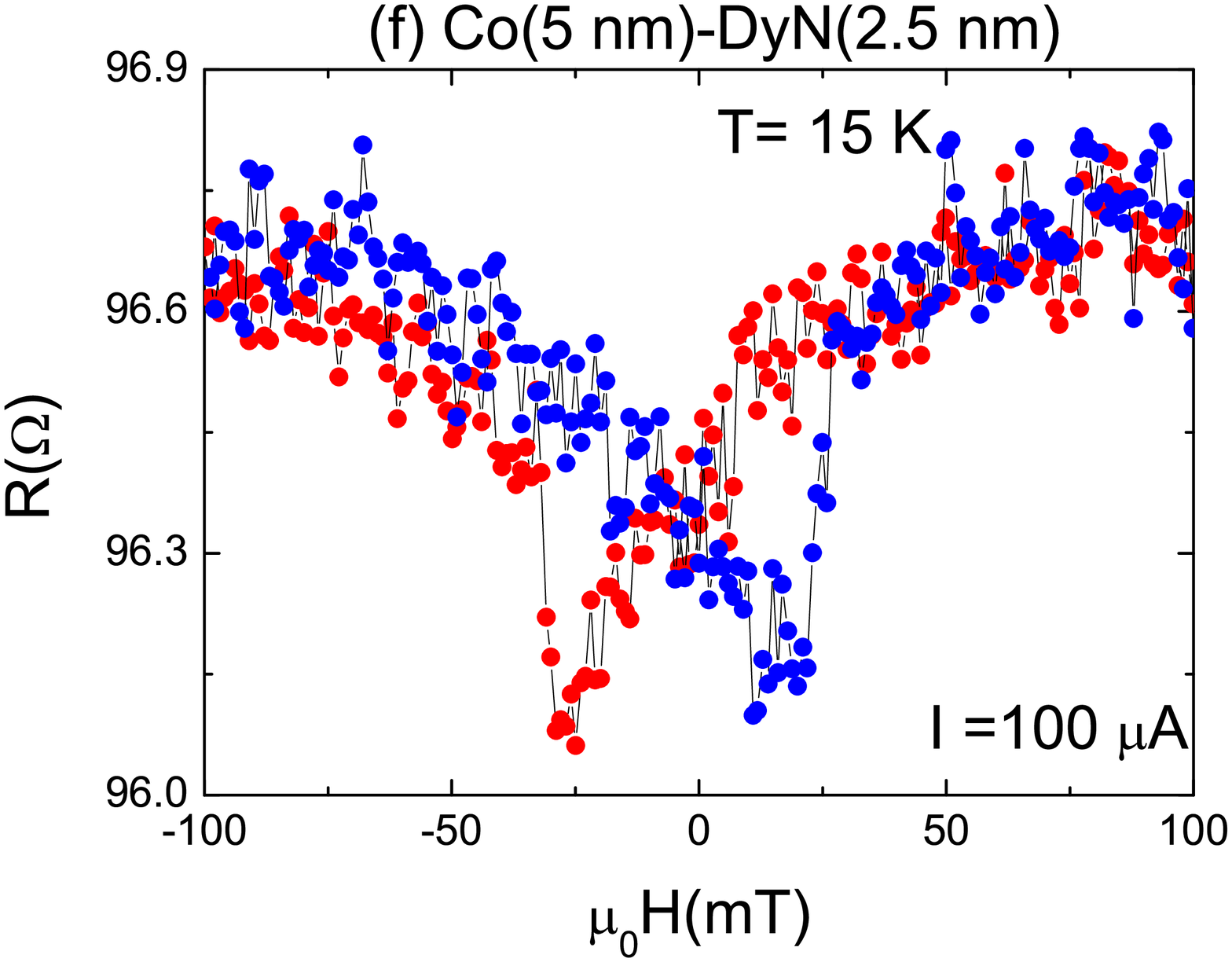}\\

  \centering
  \includegraphics[width= 6 cm]{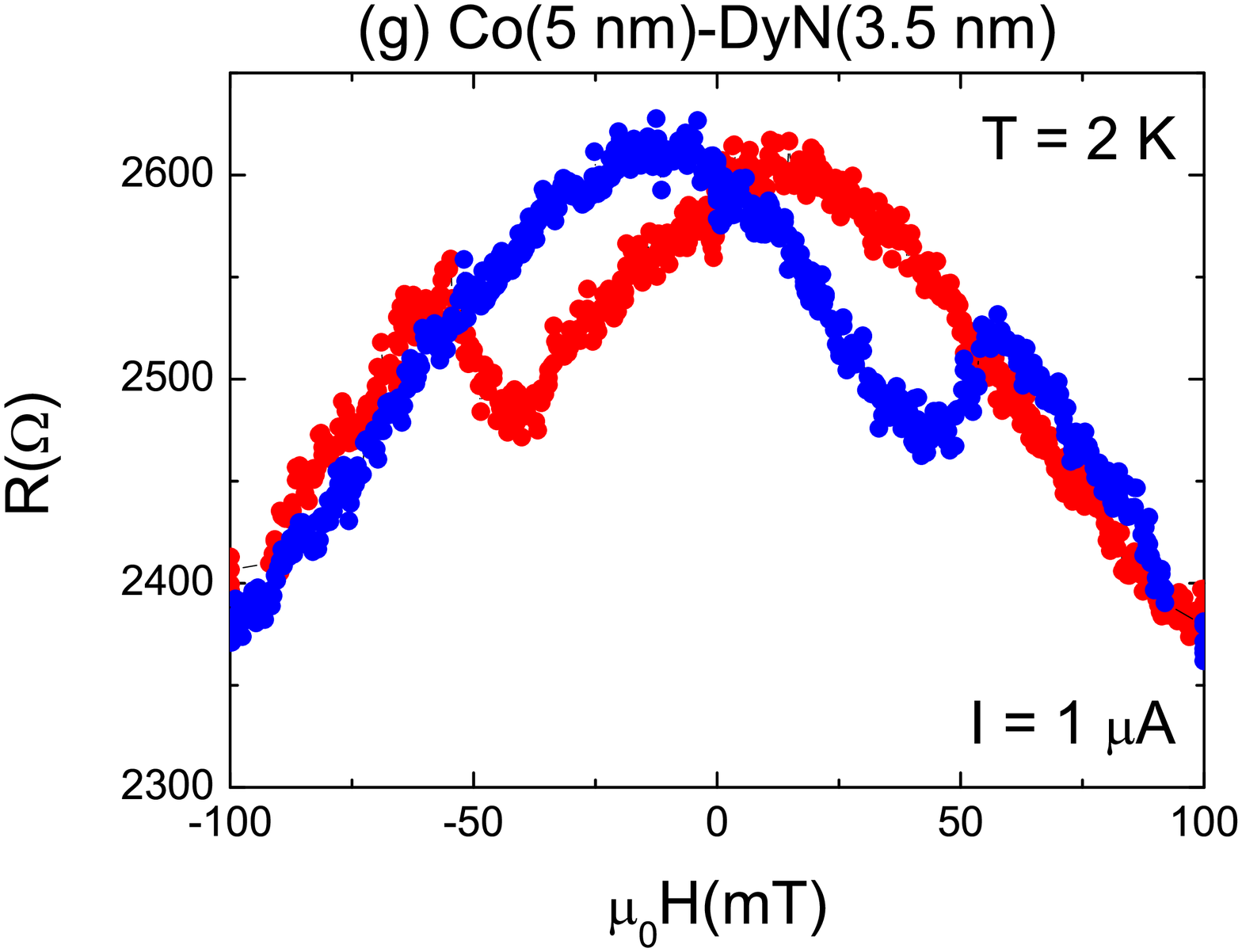}
&

  \includegraphics[width= 6 cm]{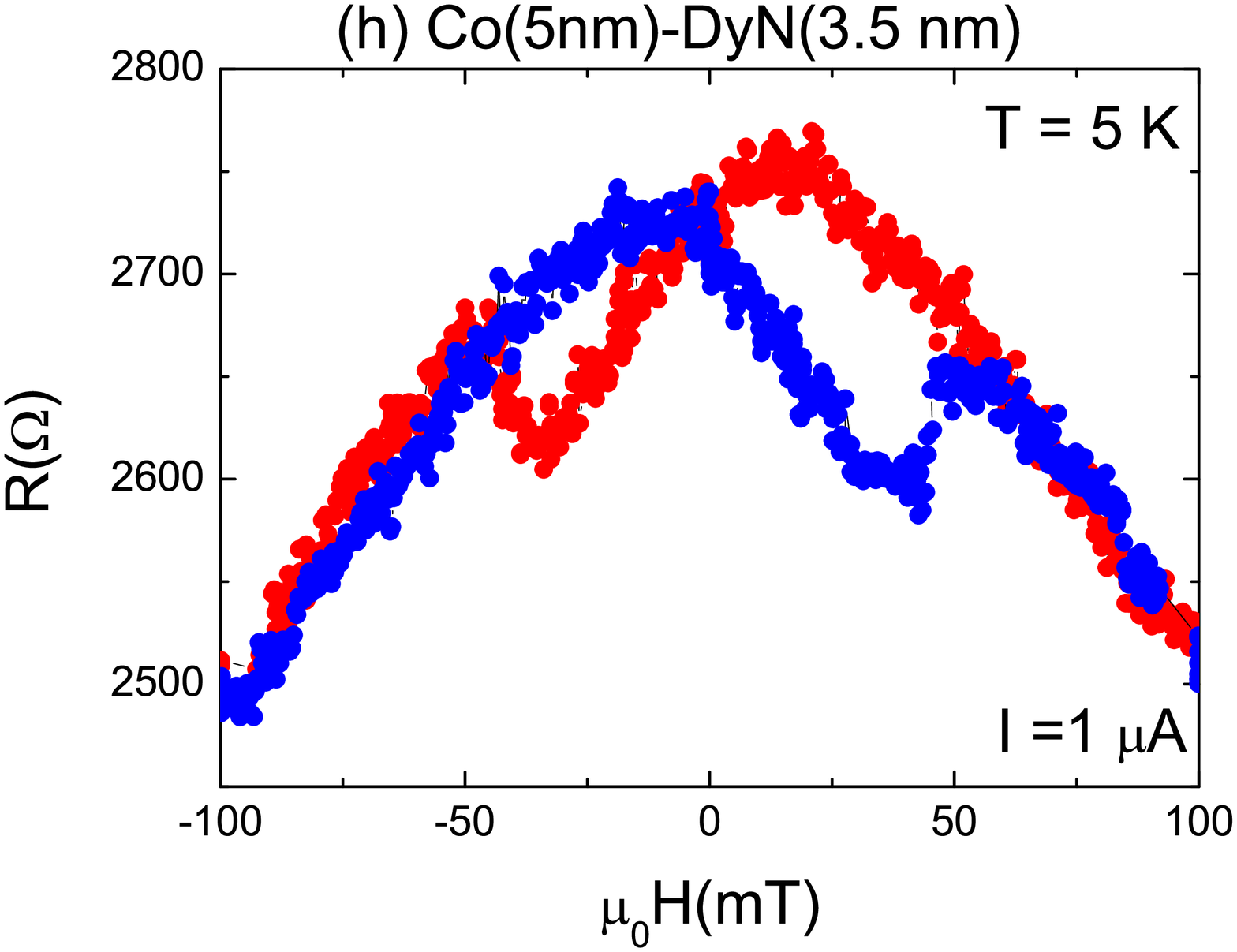}
&

  \includegraphics[width= 6 cm]{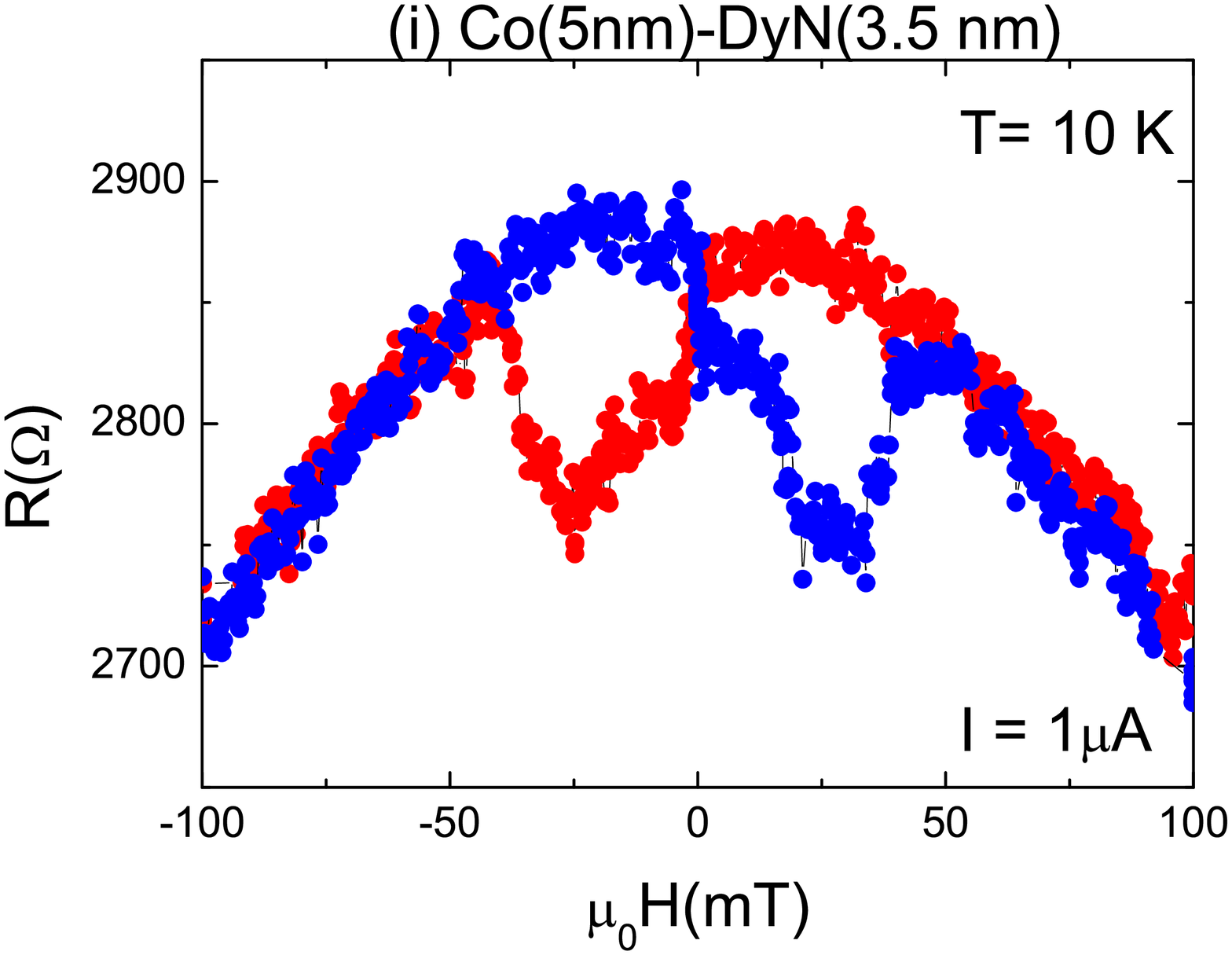}\\

 \centering
  \includegraphics[width= 6 cm]{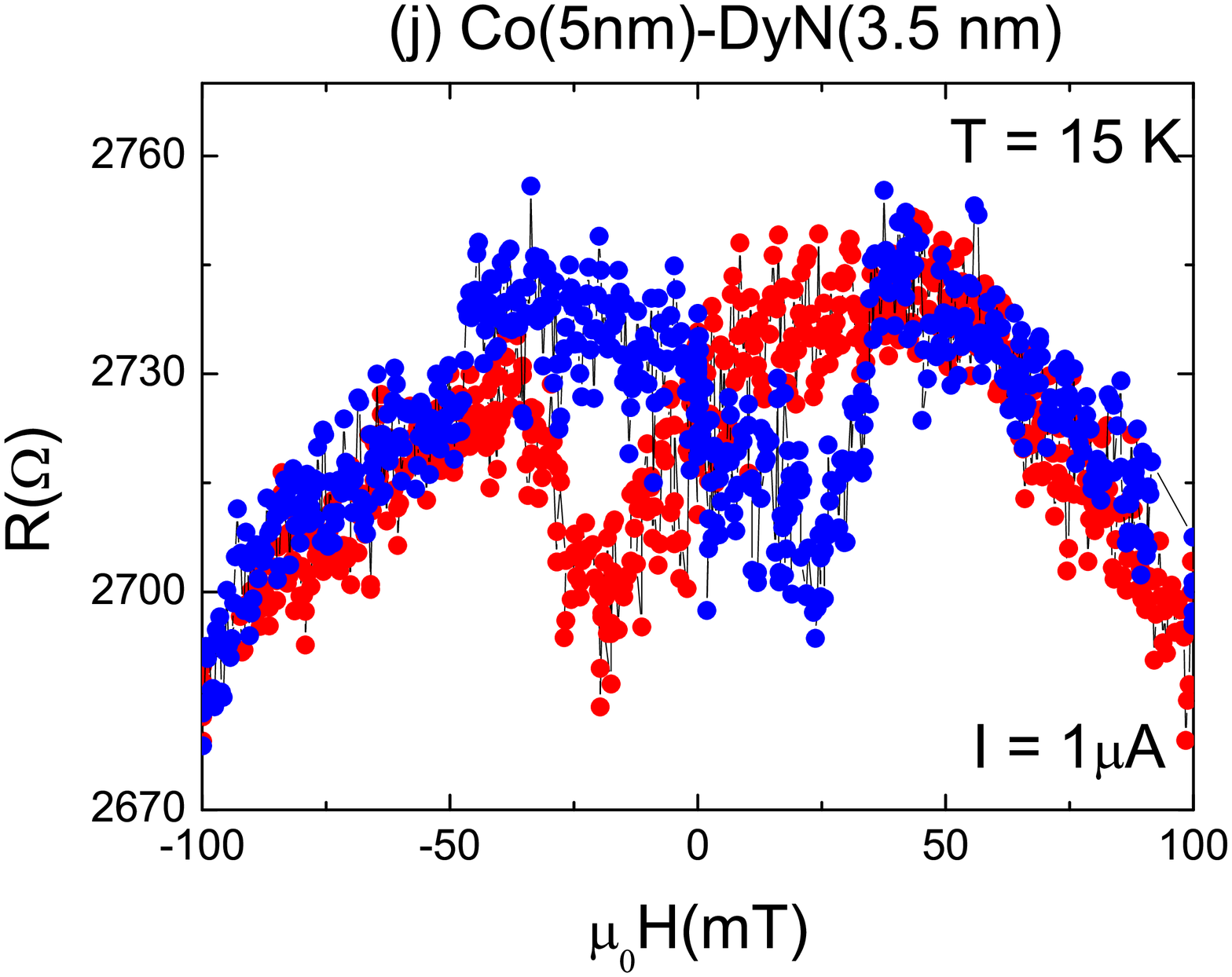}

\end{tabular}
\caption{The RH loop of the (a)-(c)Co(5 nm)-DyN(2 nm),(d)-(f)Co(5
nm)-DyN(2.5 nm) and (g)-(j) Co(5 nm)-DyN(3.5 nm) devices measured
at different temperature. The resistance switching field can be
seen between $\pm$20-30 mT. The variation in the switching field
for different thickness of DyN is probably due to be different
nitridation of Co.} \label{fig:fig}
\end{figure}



\begin{figure}[!h]
\begin{tabular}{lll}
  \centering
  \includegraphics[width= 6 cm]{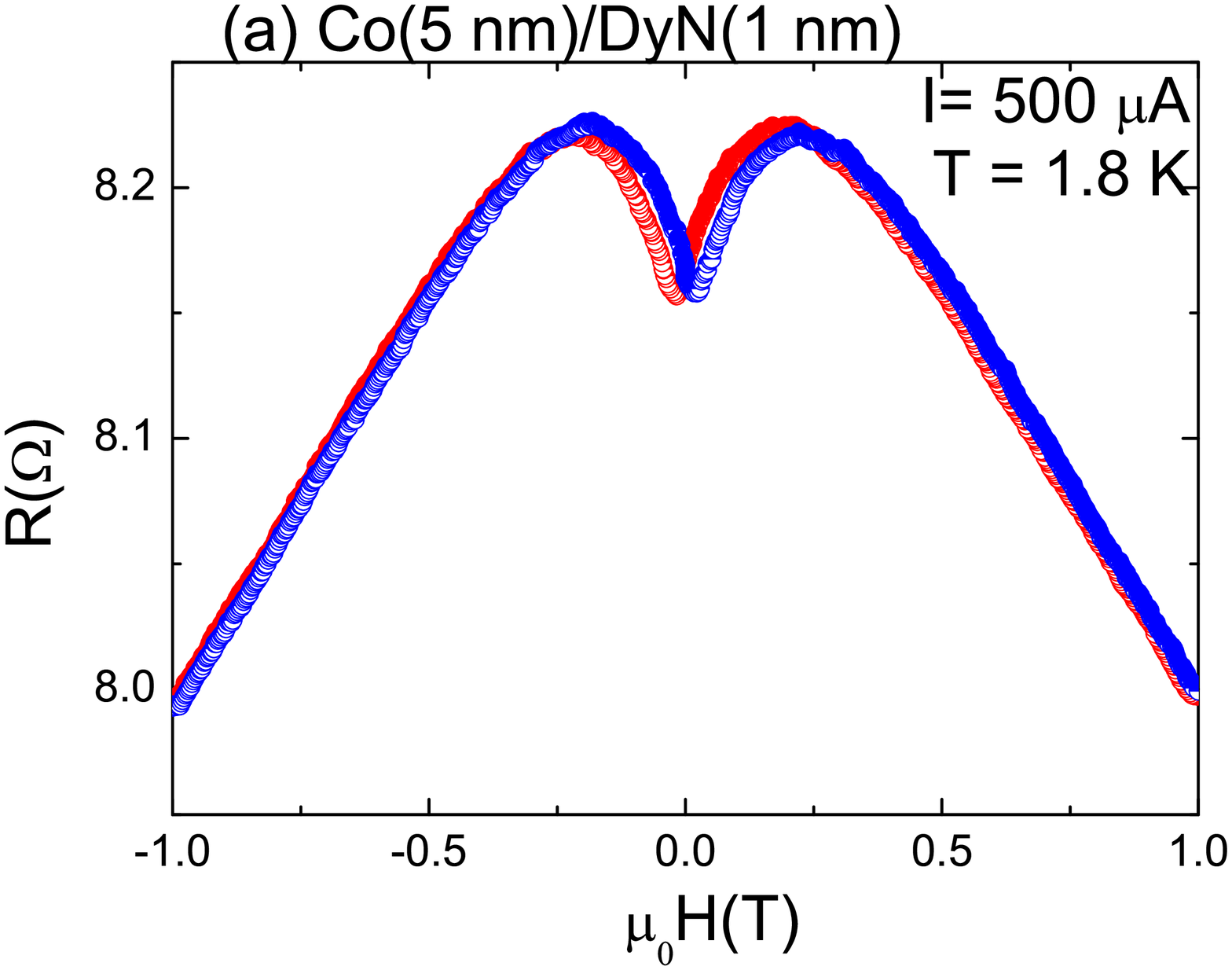}
&
  \centering
  \includegraphics[width= 6 cm]{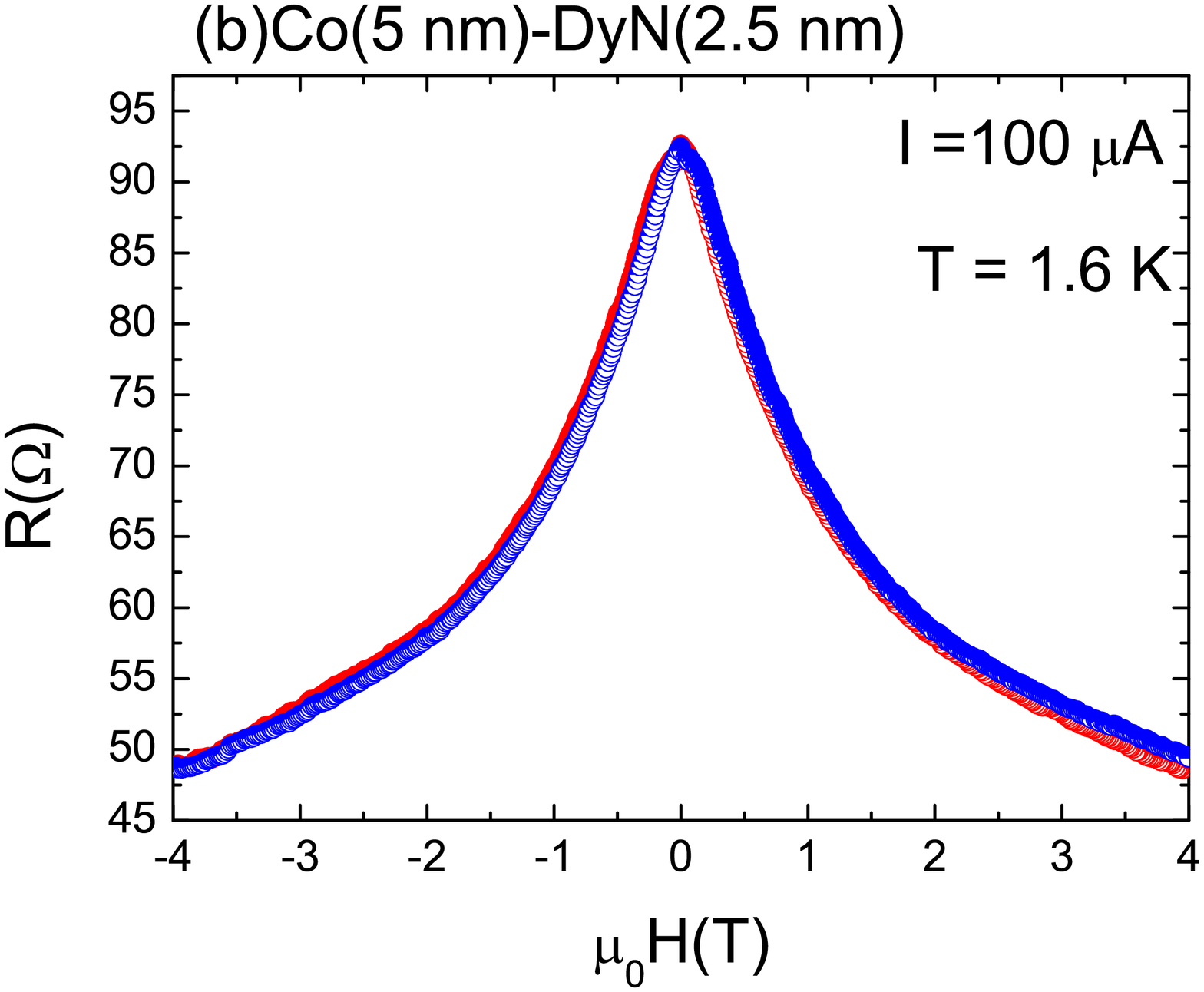}
&
 \centering
 \includegraphics[width= 6 cm]{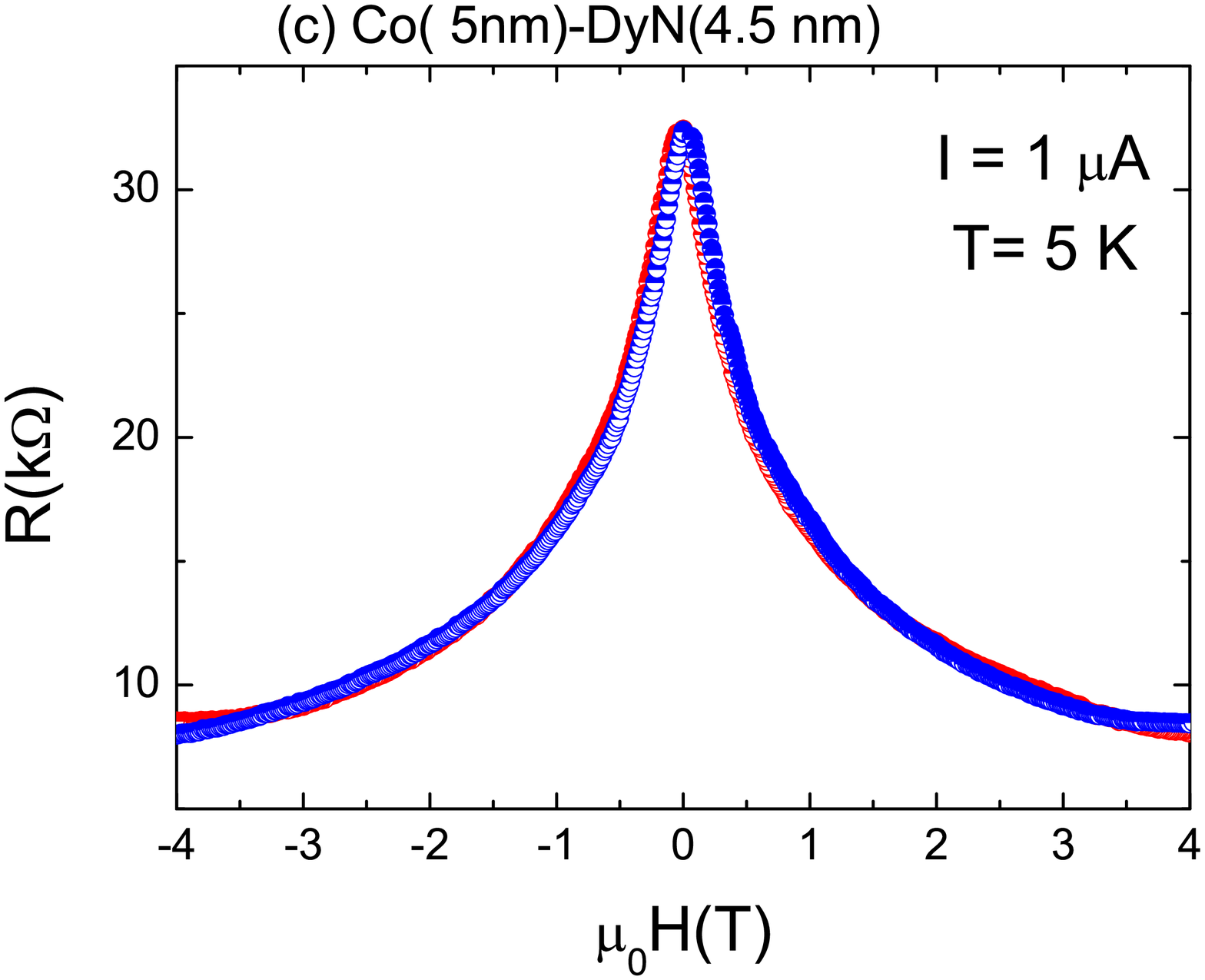}
\end{tabular}
\caption{The RH loop of the Co(5 nm)-DyN(\emph{t} nm) device
measured up to high field $\mu_0H$ = 4 T. }
\end{figure}

\begin{figure}[!h]
\begin{tabular}{lll}
  \centering
  \includegraphics[width= 6 cm]{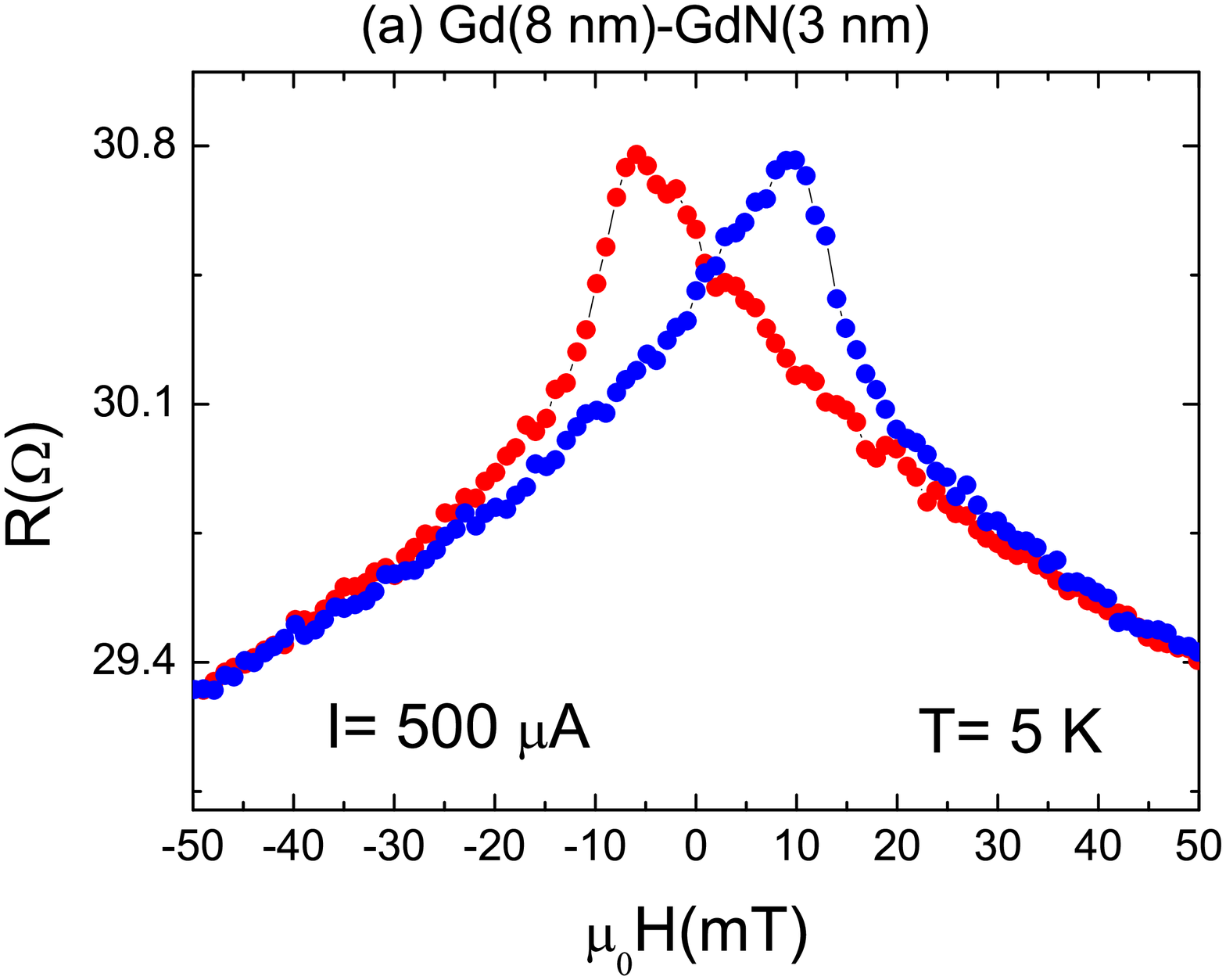}
&
  \centering
  \includegraphics[width= 6 cm]{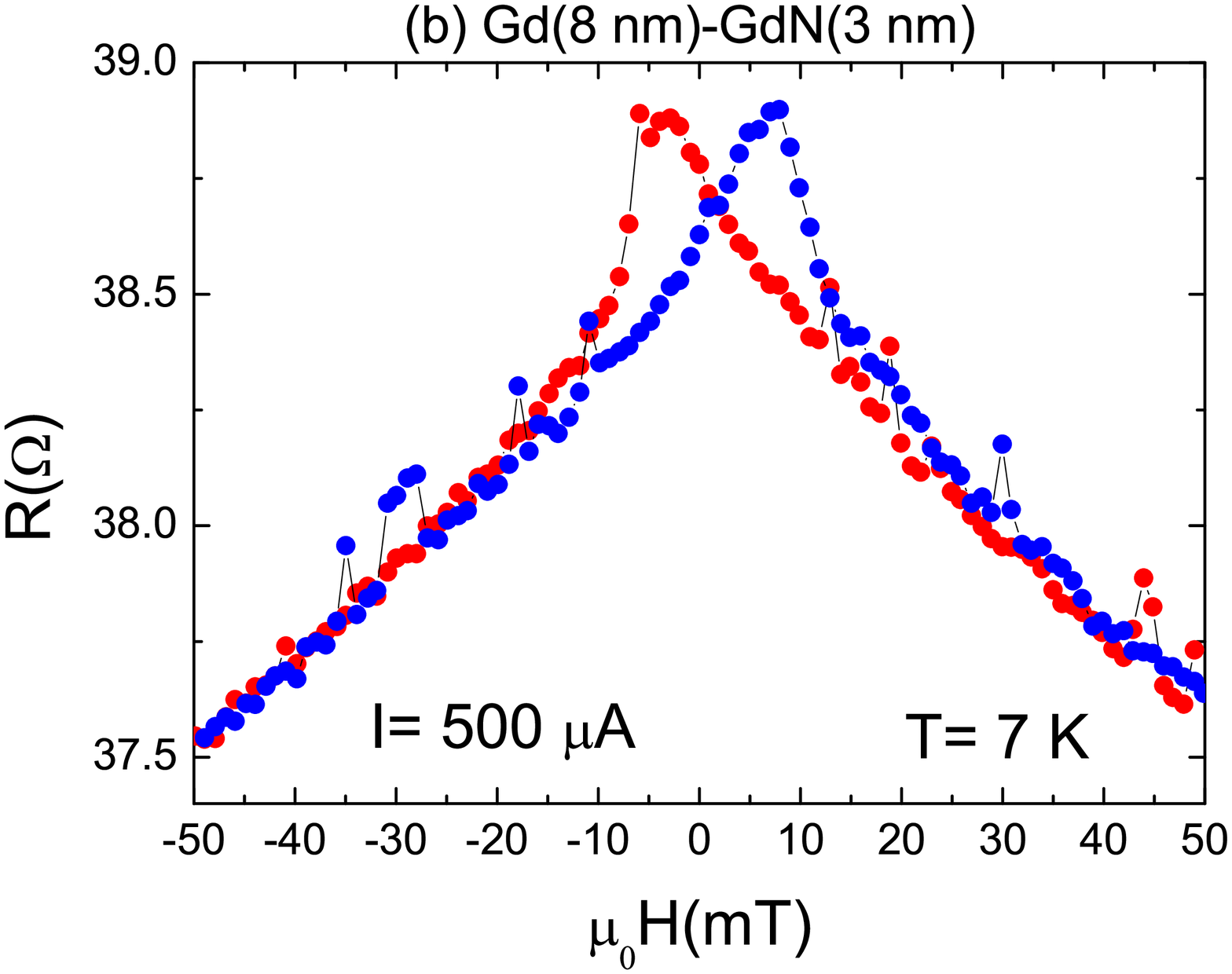}
&
 \centering
 \includegraphics[width= 6 cm]{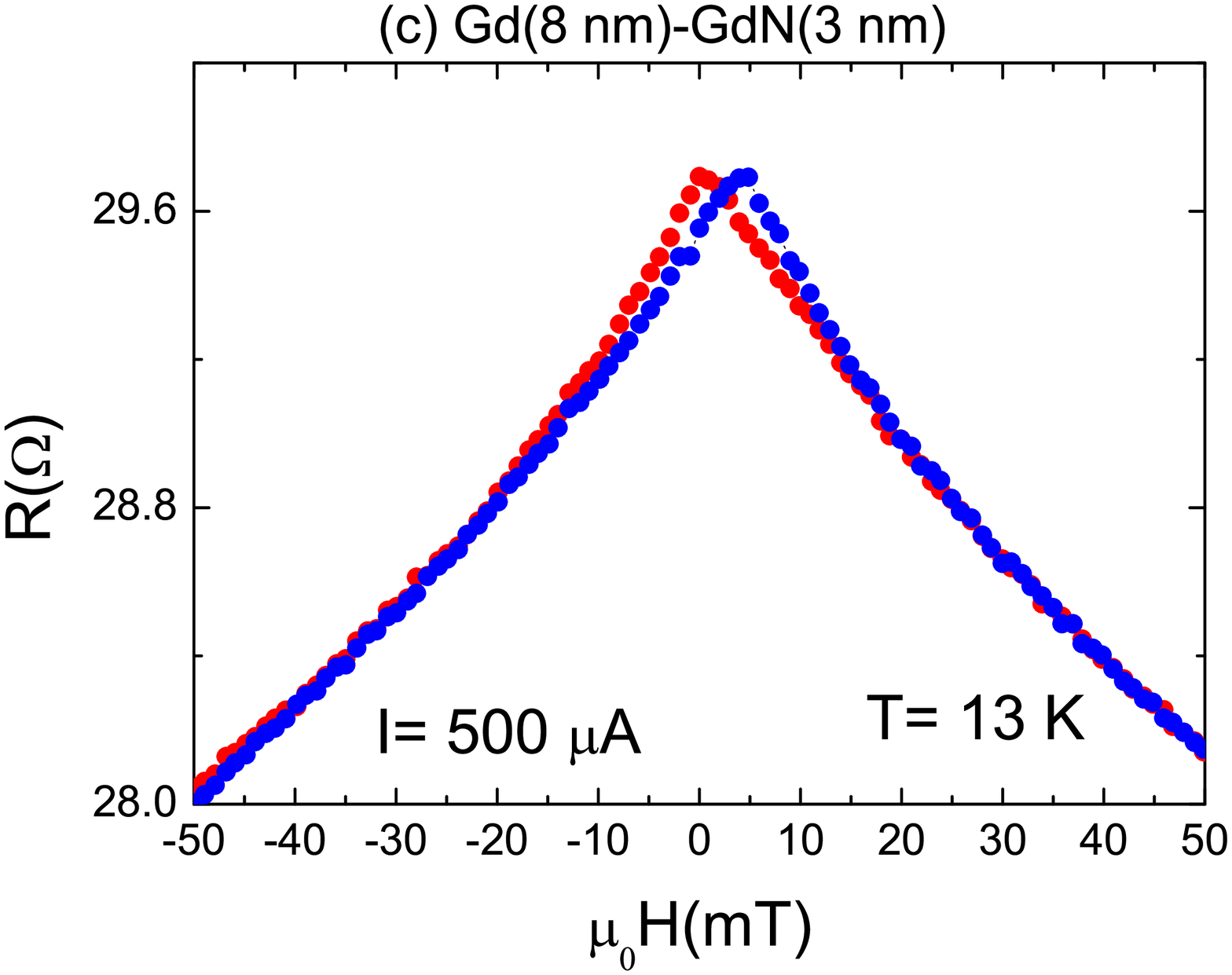}
\end{tabular}
\caption{R-H loop of the Gd(8 nm)-GdN(3 nm) device measured at
different temperature with bias current $I = 500$ $\mu$A.}
\end{figure}

\begin{figure}[!h]
\begin{tabular}{ll}
  \centering
  \includegraphics[width= 6 cm]{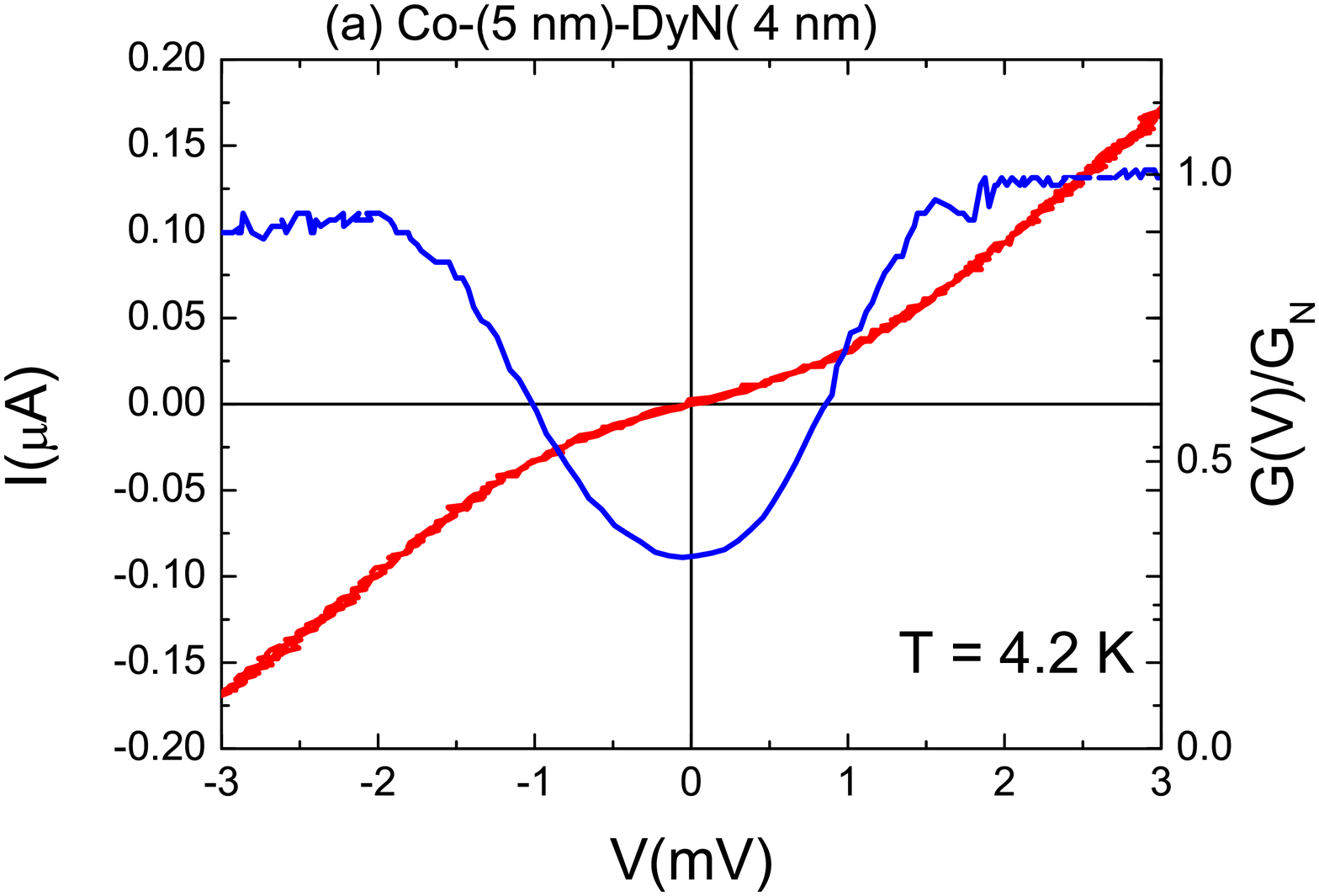}
&

  \includegraphics[width= 6 cm]{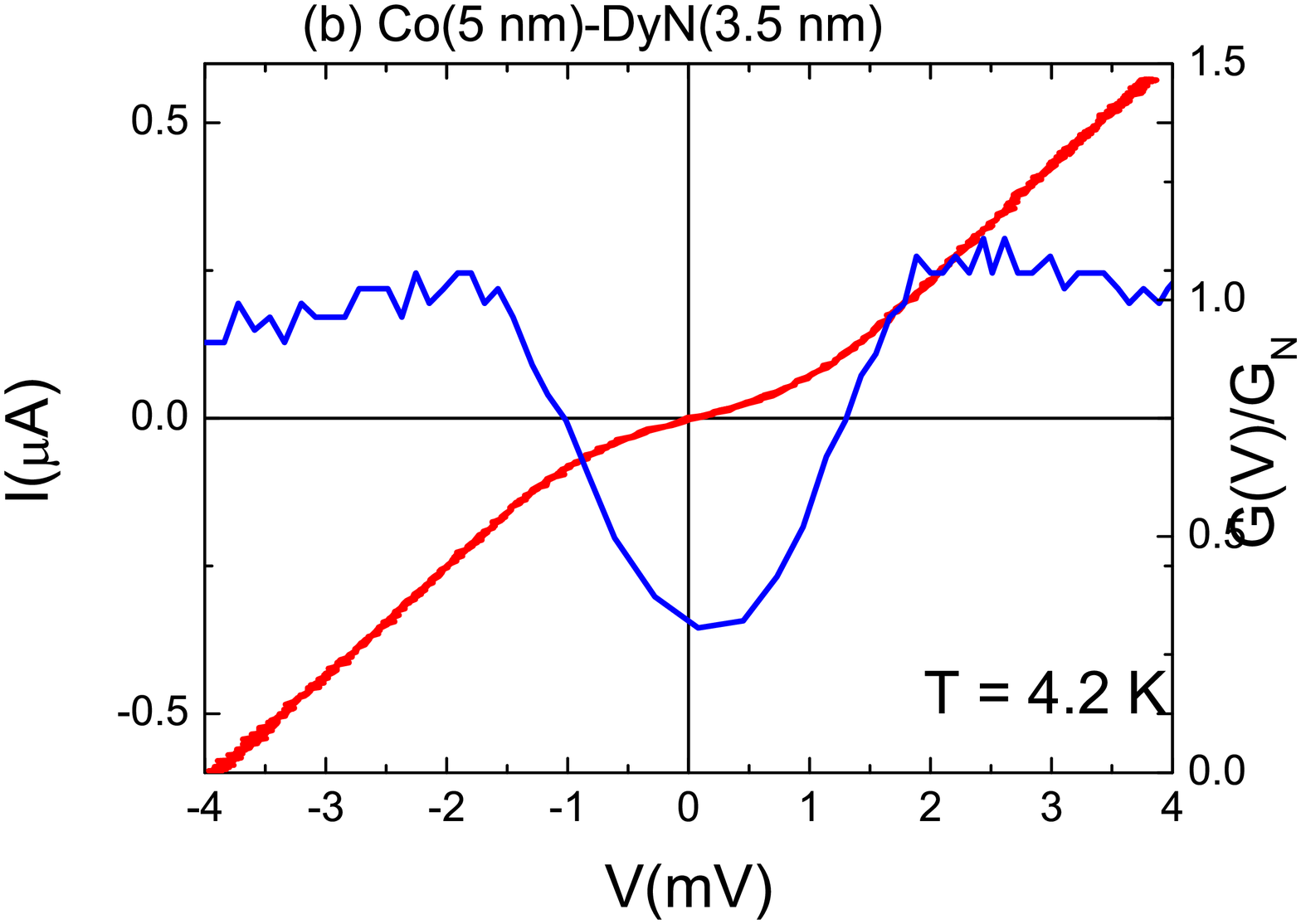}\\

 \centering
 \includegraphics[width= 6 cm]{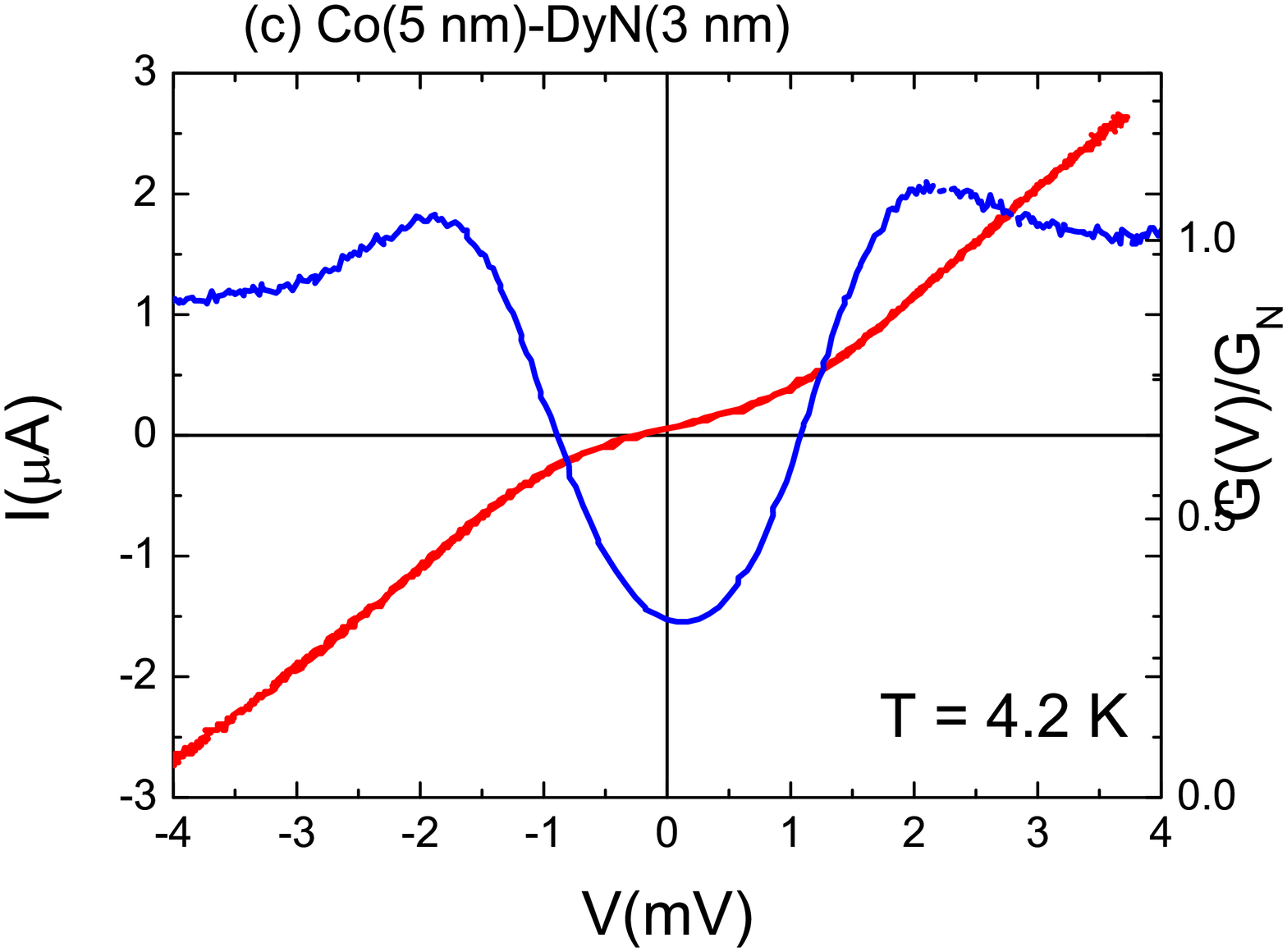}
&
 \includegraphics[width= 6 cm]{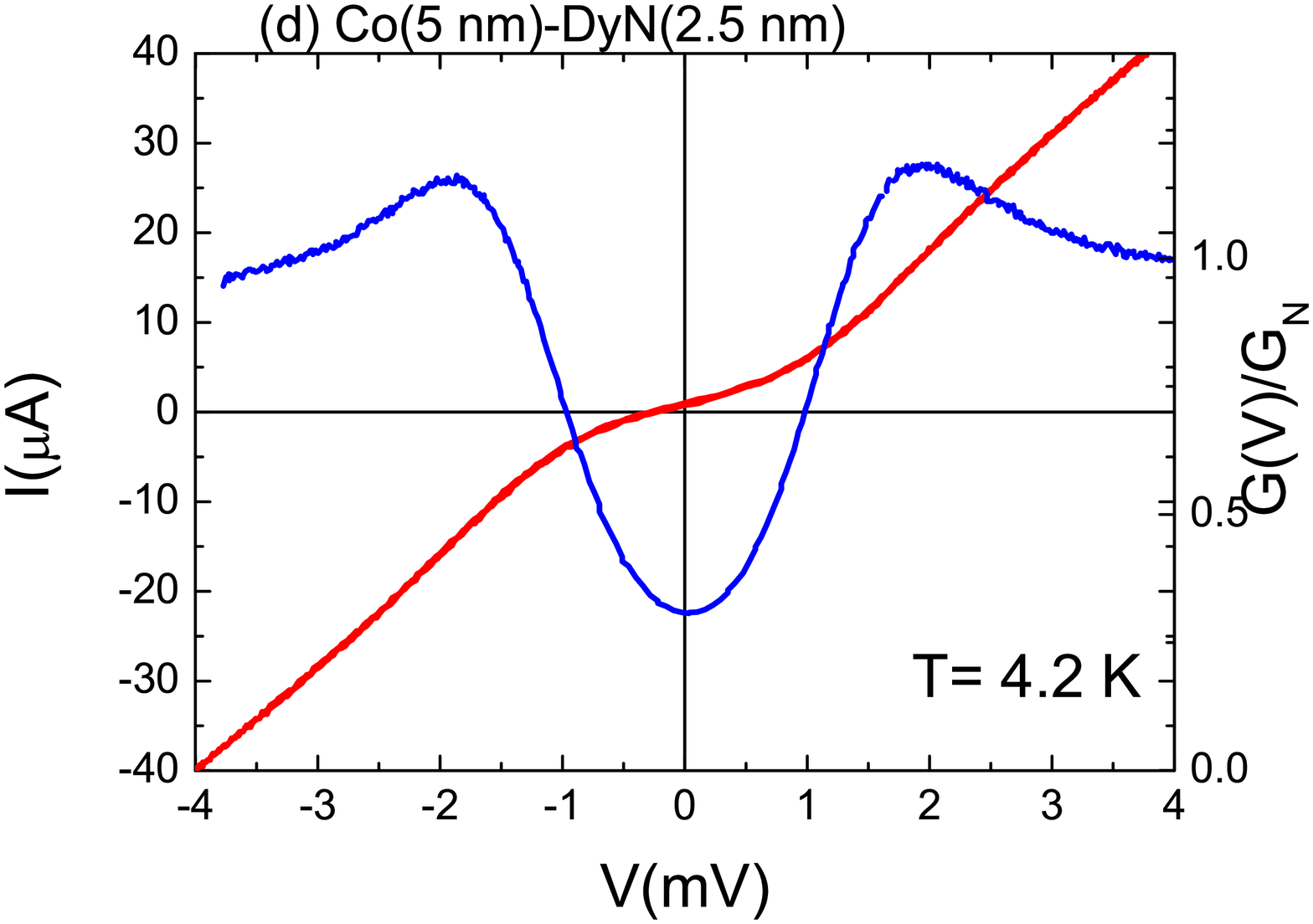}\\

 \centering
 \includegraphics[width= 6 cm]{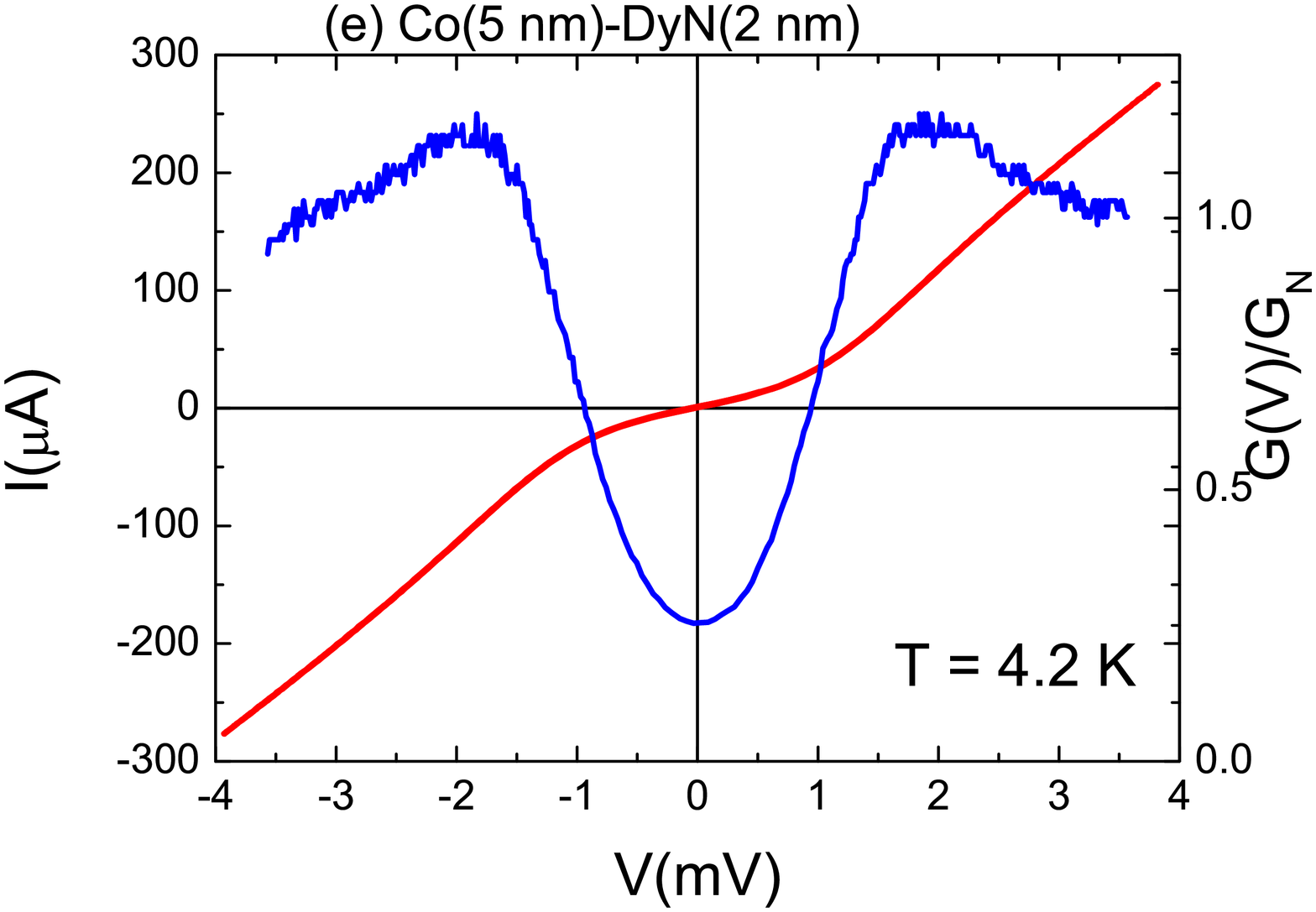}
&

 \includegraphics[width= 6 cm]{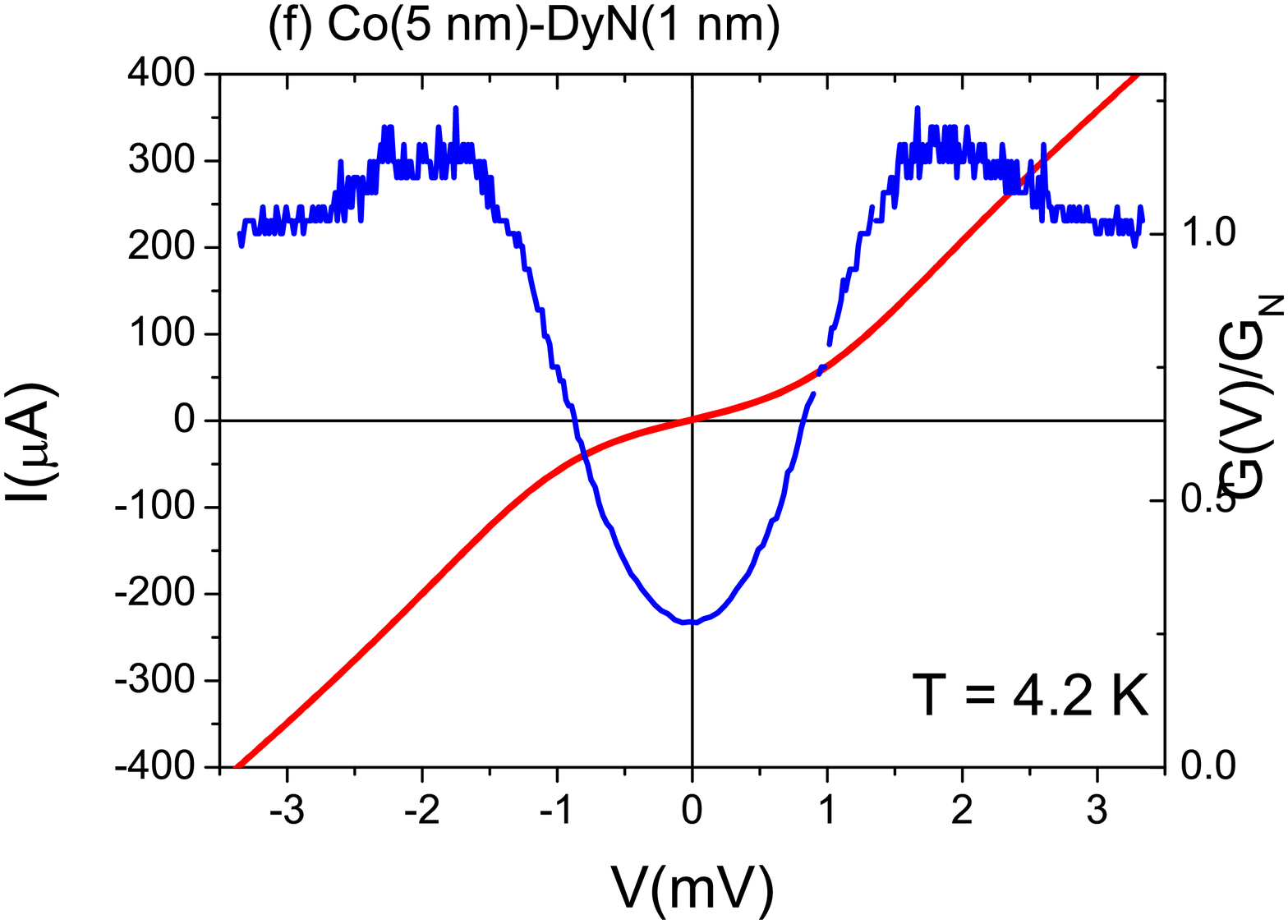}
\end{tabular}
\caption{I-V and normalized conductance spectra ($G(V)/G_N$) of
Co-DyN devices with different thickness of DyN.}
\end{figure}

\begin{figure}
\begin{tabular}{ll}
  \centering
  \includegraphics[width= 6 cm]{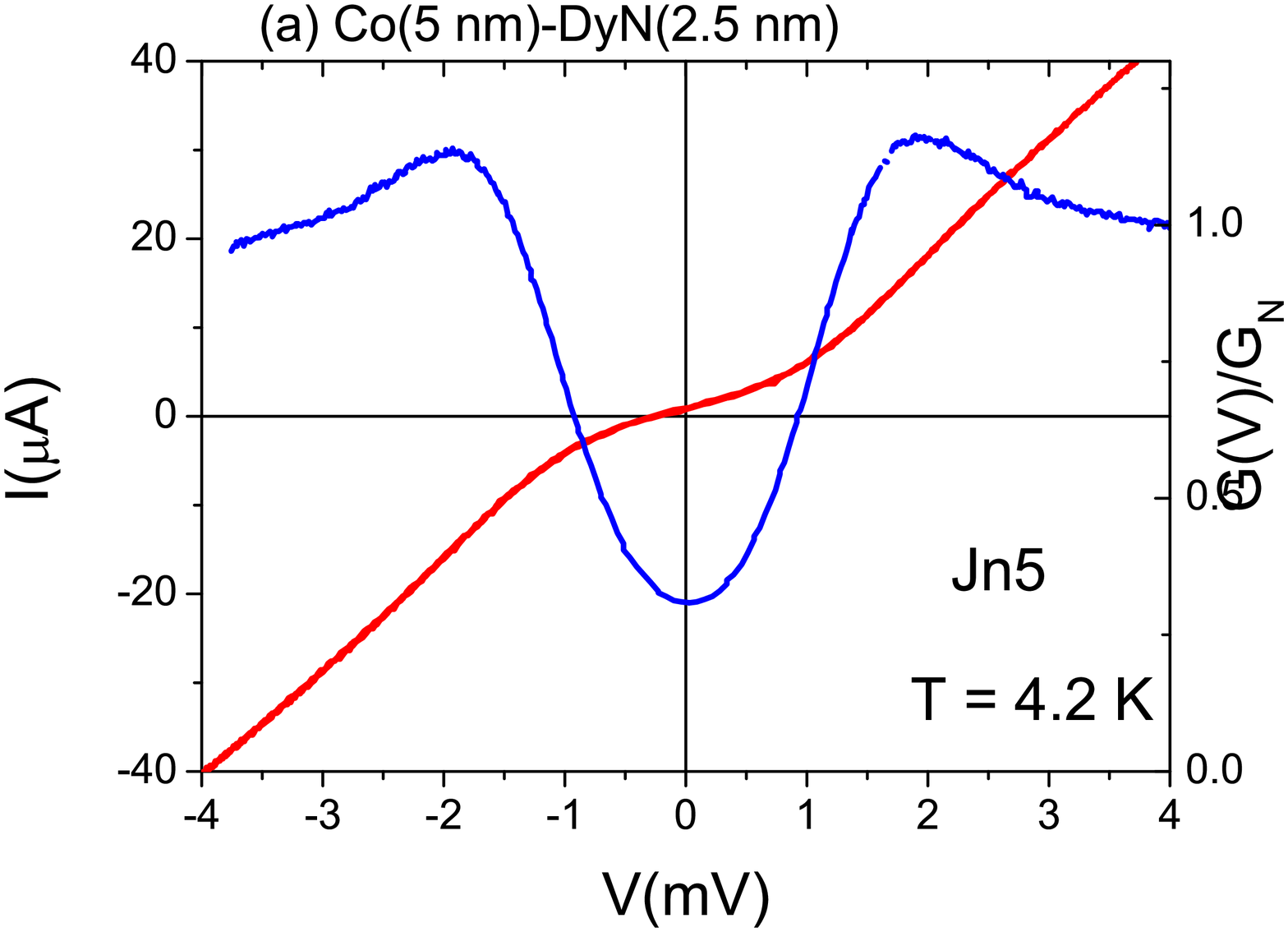}
&

  \includegraphics[width= 6 cm]{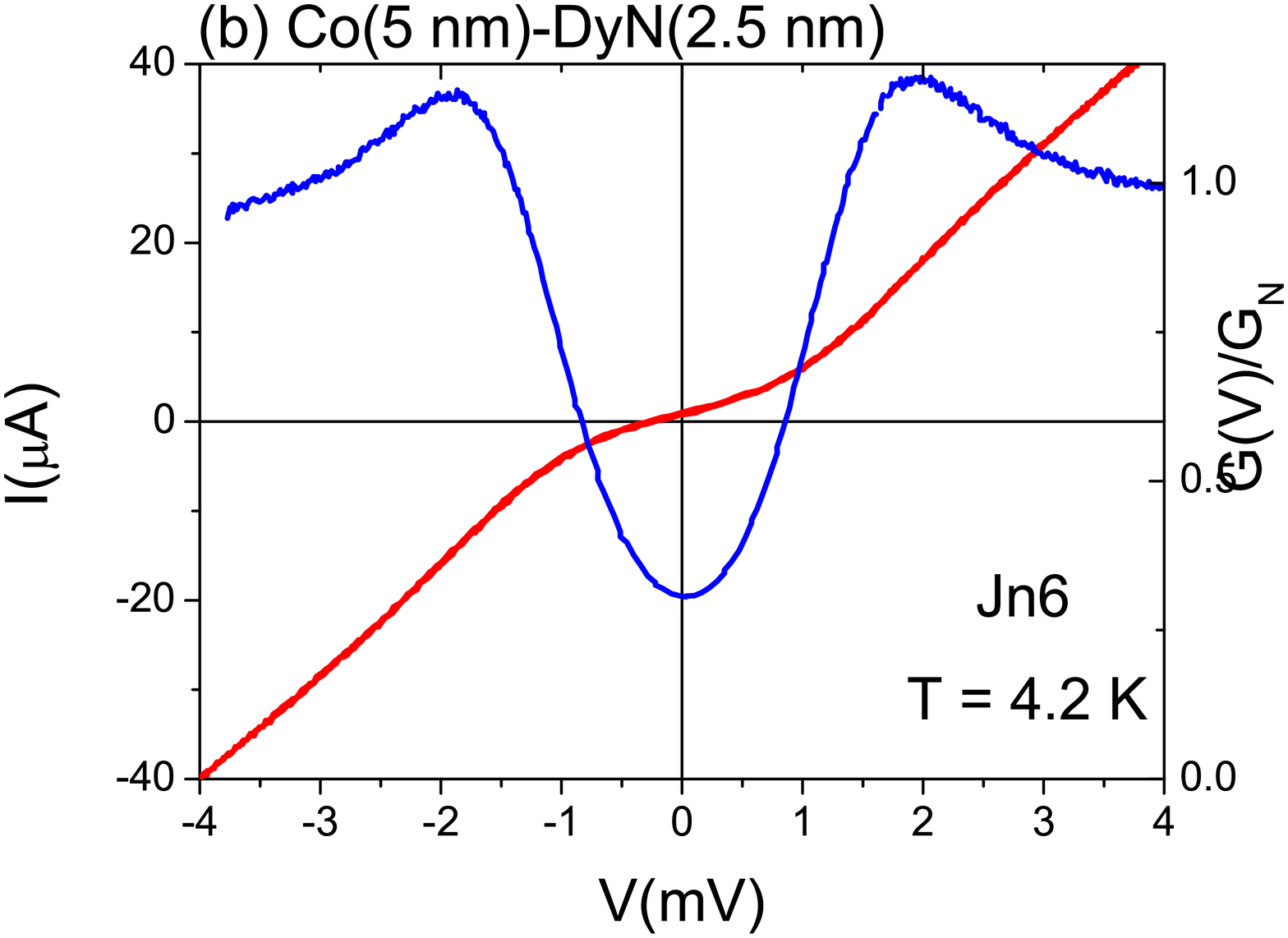}\\

 \centering
 \includegraphics[width= 6 cm]{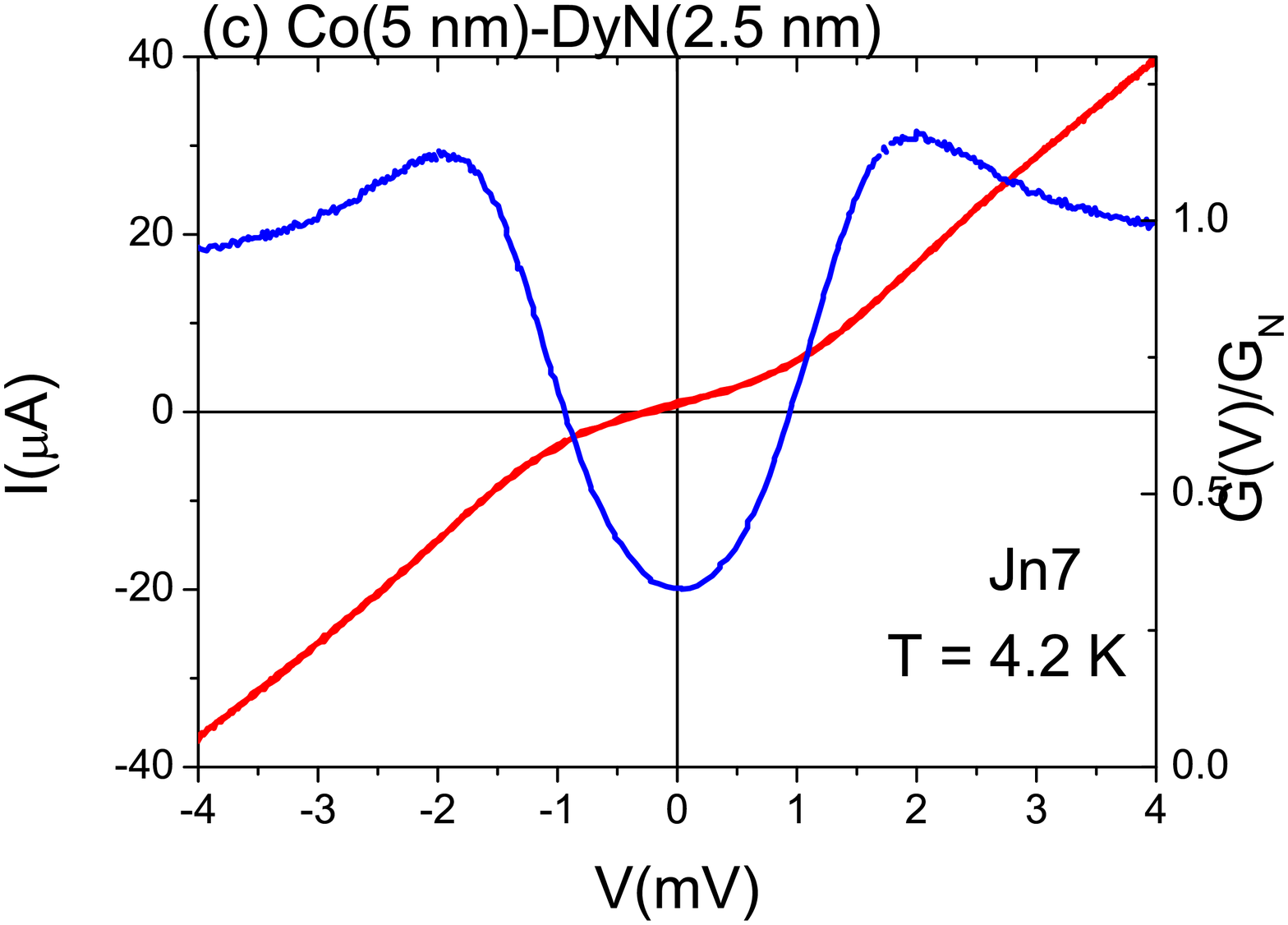}
&
 \includegraphics[width= 6 cm]{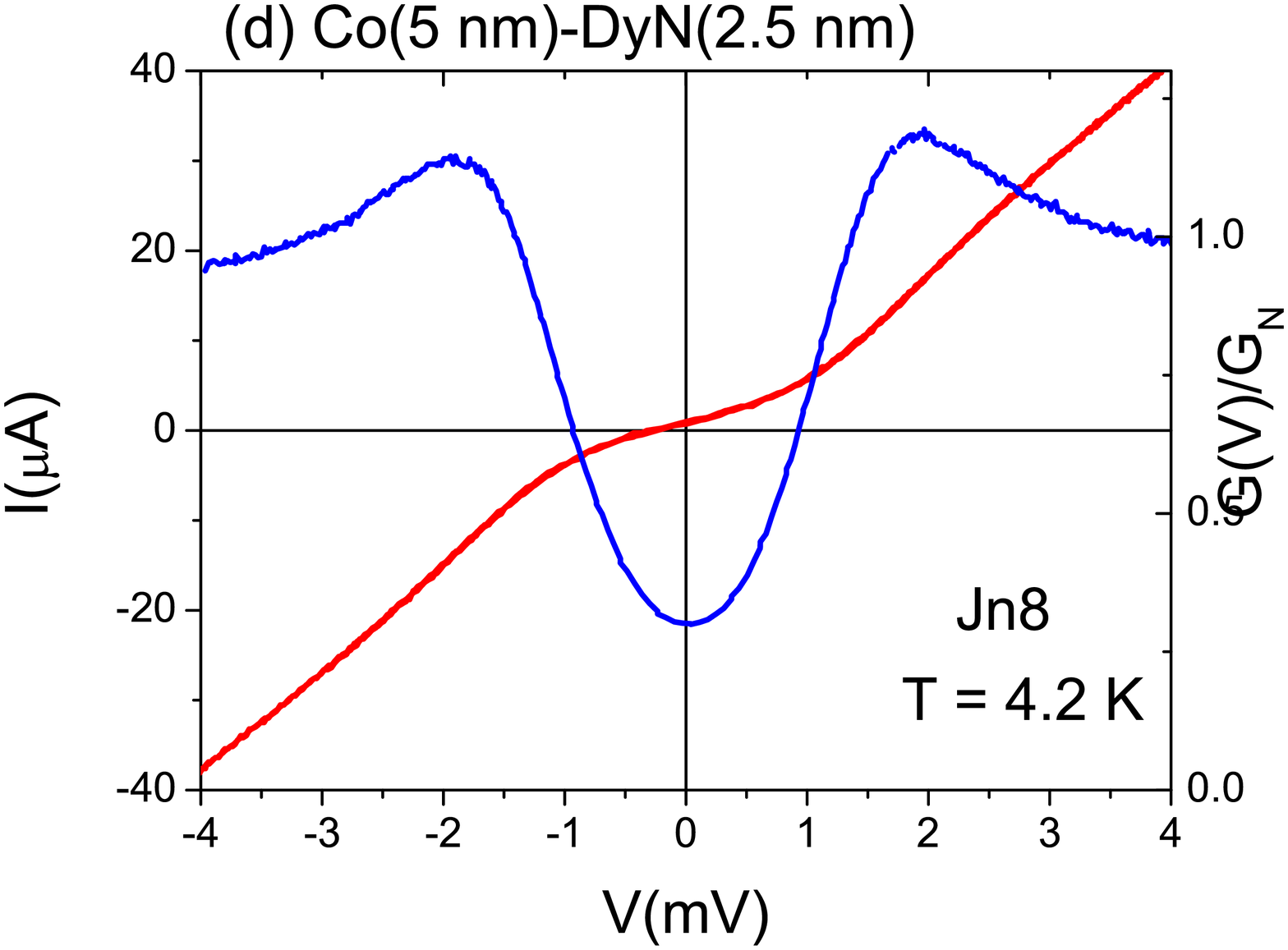}

\end{tabular}
\caption{I-V and normalized conductance spectra ($G(V)/G_N$) of
different Co-DyN device on the same chip. Each chip contain 8
identical junctions. This shows the reproducibility from device to
device is quite good.}
\end{figure}

\begin{figure}[!h]
\begin{tabular}{ll}
  \centering
  \includegraphics[width= 6 cm]{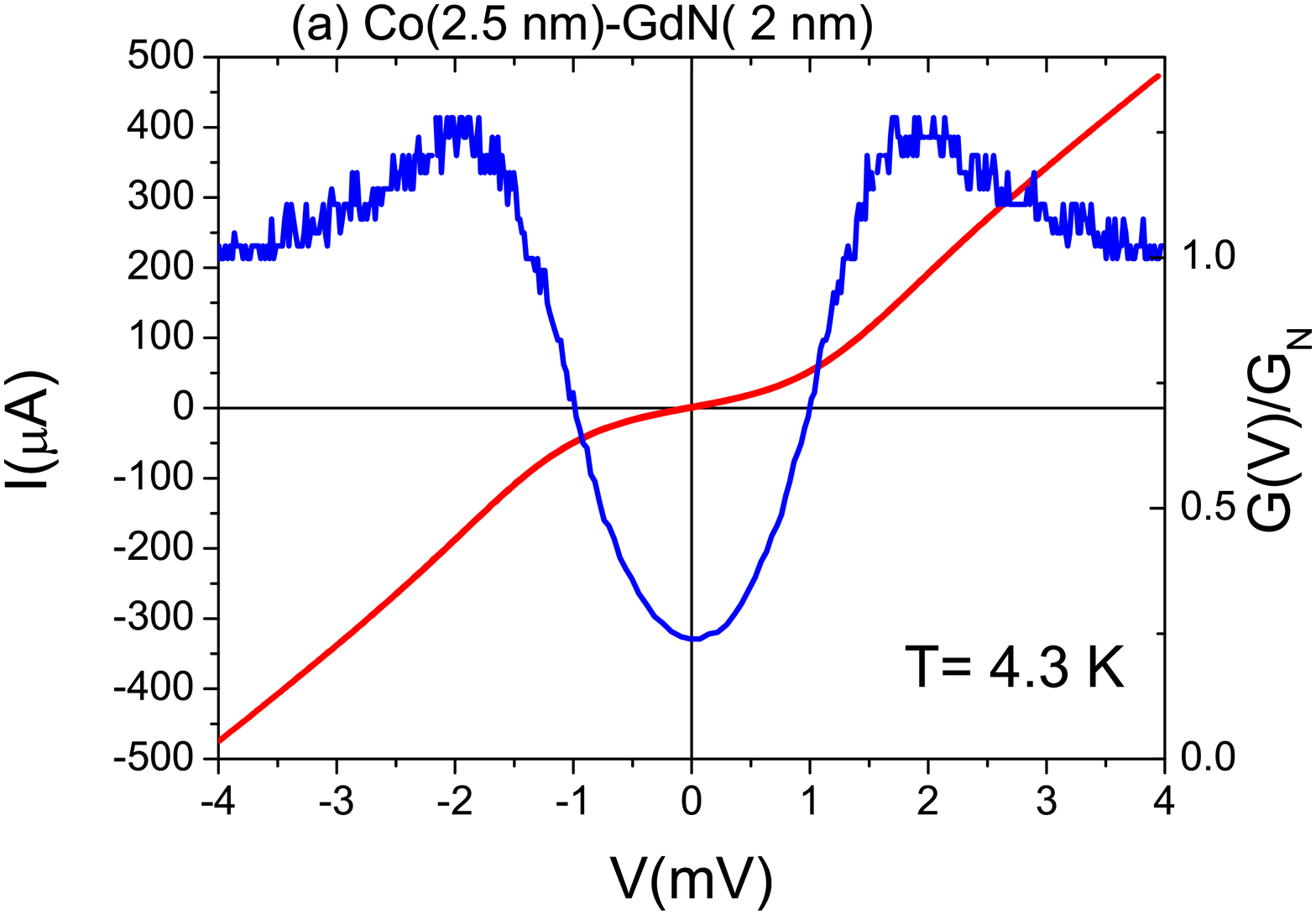}
&
  \centering
  \includegraphics[width= 6 cm]{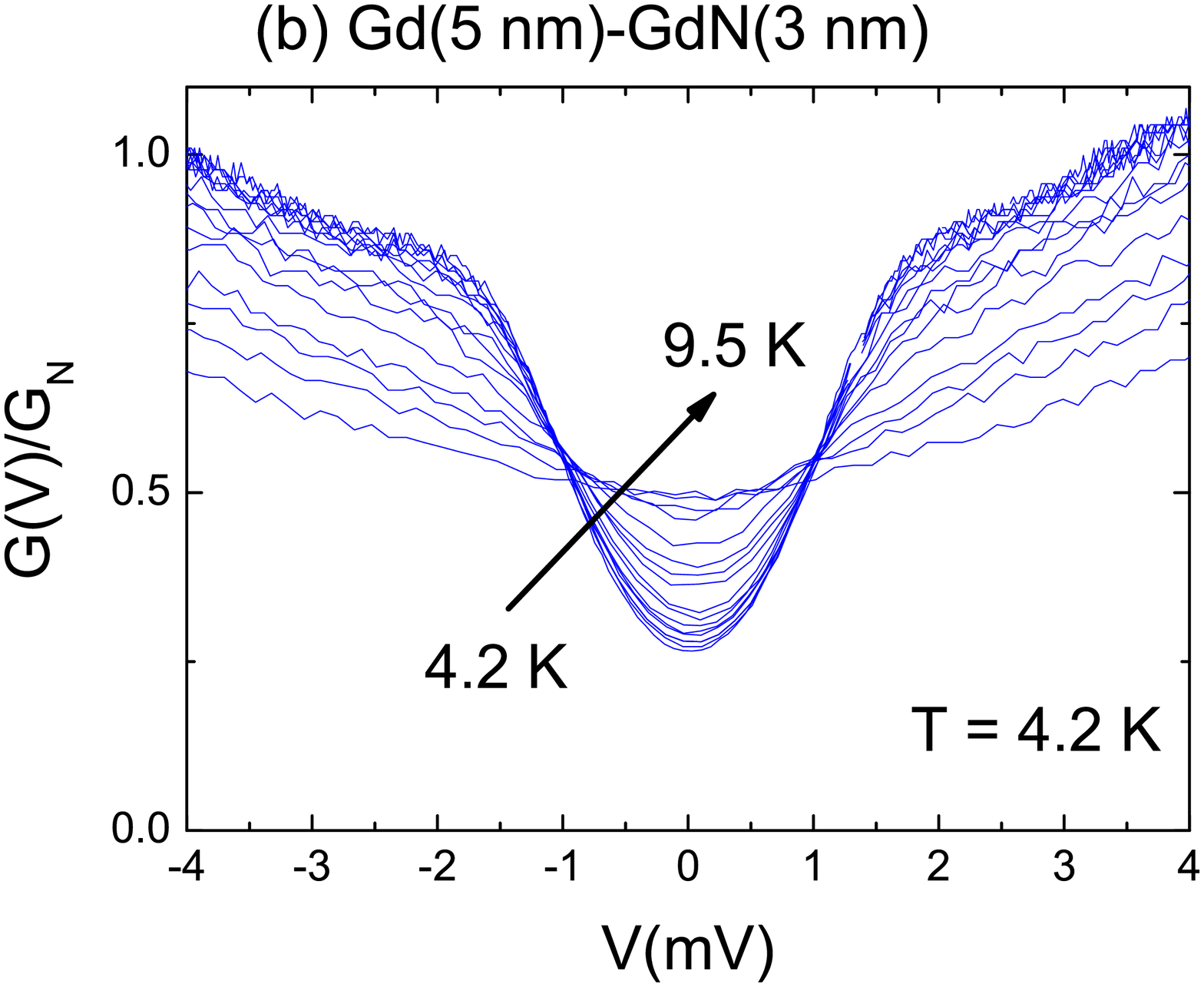}

\end{tabular}
\caption{I-V and normalized conductance spectra $G(V)/G_N$ of (a)
Co-GdN and (b) Gd-GdN device device (The conductance spectra at
different temperature is shifted lightly for clarity).}
\end{figure}

\end{widetext}

\end{document}